\documentclass[11pt]{article}
\usepackage{amsmath,amssymb,color,graphics,epsfig,cite}

\usepackage{amsfonts}

\newcommand{\be}{\begin{equation}}
\newcommand{\ee}{\end{equation}}
\newcommand{\bea}{\setlength\arraycolsep{2pt} \begin{eqnarray}}
\newcommand{\eea}{\end{eqnarray}}

\def\0{{\sst{(0)}}}
\def\1{{\sst{(1)}}}
\def\2{{\sst{(2)}}}
\def\3{{\sst{(3)}}}
\def\4{{\sst{(4)}}}
\def\5{{\sst{(5)}}}
\def\6{{\sst{(6)}}}
\def\7{{\sst{(7)}}}
\def\8{{\sst{(8)}}}
\def\sst#1{{\scriptscriptstyle #1}}
\def\oneone{\rlap 1\mkern4mu{\rm l}}

\begin{document}

\begin{center}
{\Large {\bf Black Hole Thermodynamic Ensembles, Euclidean Action and Legendre Transformation }}

\vspace{20pt}

Liang Ma

\vspace{10pt}

{\it Center for Joint Quantum Studies, Department of Physics,\\
School of Science, Tianjin University, Tianjin 300350, China }

\vspace{40pt}

\underline{ABSTRACT}
\end{center}

In thermodynamics, a Legendre transformation of the free energy provides a mapping between different statistical ensembles. In this work, we demonstrate that performing a Legendre transformation of the black hole on-shell action is equivalent to imposing different boundary conditions on the fields. Consequently, the choice of ensemble must be consistent with, and cannot contradict, the imposed boundary conditions. From this perspective, it follows that for four-dimensional dyonic black holes, the on-shell action can only be expressed either as a function of the electric charge and the magnetic potential, or alternatively as a function of the magnetic charge and the electric potential. Inspired by the Legendre transformation of the Maxwell field, we argue that for purely gravitational theories whose metric geometries admit a \(U(1)\) fiber bundle structure, i.e.\ rotating, boosted, or Kaluza-Klein monopole configurations, one can similarly introduce appropriate Legendre terms, in the sense of dimensional reduction, to modify the thermodynamic ensemble of the black hole. Within the dimensional reduction framework, we study the on-shell action of black holes in five-dimensional minimal supergravity with a Chern-Simons term, analyze the corresponding Legendre transformation procedure, and show how the resulting formulation remains consistent with the Wald formalism.

\vfill {\footnotesize maliang0@tju.edu.cn}


\thispagestyle{empty}
\pagebreak

\tableofcontents
\addtocontents{toc}{\protect\setcounter{tocdepth}{2}}

\newpage

\section{Introduction}

Since the seminal works of Bekenstein and Hawking established the foundations of black hole thermodynamics \cite{Bekenstein:1973ur,Hawking:1975vcx}, black holes have been extensively investigated as genuine thermodynamic systems. Subsequently, Gibbons and Hawking showed that the Euclidean on-shell action of a black hole is identified with its thermodynamic free energy \cite{Gibbons:1976ue}, thereby reducing the analysis to a conventional thermodynamic computation.
 To evaluate the gravitational Euclidean on-shell action properly, one must add the Gibbons-Hawking-York boundary term \cite{Gibbons:1976ue,York:1972sj} to impose suitable boundary conditions, as well as a background subtraction term to cancel divergences. For AdS black holes, these divergences can alternatively be removed using boundary counterterms \cite{Balasubramanian:1999re}, which is a technique known as holographic renormalization \cite{deHaro:2000vlm,Bianchi:2001kw,Skenderis:2002wp} within the framework of AdS/CFT \cite{Maldacena:1997re}. Once the thermodynamic free energy is known, the full set of black hole thermodynamic quantities can be unambiguously extracted using standard thermodynamic relations, in complete agreement with the first law of black hole thermodynamics. In this respect, black hole thermodynamics is fundamentally no different from the thermodynamics of ordinary physical systems \cite{Almheiri:2020cfm}.

In thermodynamics and statistical mechanics, different ensembles are defined based on which physical quantities are held fixed. The thermodynamic properties of a system can vary significantly depending on the chosen ensemble. We can transite between different ensembles via Legendre transformations. From the perspective of the variational principle in classical mechanics, a Legendre transformation effectively modifies the system's boundary conditions by adding an appropriate total derivative term. Therefore, the validity of a Legendre transformation must be consistent with the variational principle. Similarly, when studying the thermodynamic stability of black holes, it is crucial to first specify the ensemble before conducting the analysis. In the case of 4D dyonic black holes, an interesting observation is that although the electric and magnetic charges enter symmetrically in the metric, this symmetry is broken at the level of the on-shell action. This inequivalence implies that some ensembles in dyonic Einstein-Maxwell theory may be inadmissible, as they conflict with the variational principle. This example illustrates a crucial distinction between the free energy defined via the on-shell action and that of conventional thermodynamics. In the former case, the free energy is fundamentally tied to the variational principle, and any Legendre transformation must be consistent with the prescribed boundary conditions. Consequently, thermodynamic ensembles that are incompatible with the chosen boundary conditions cannot be consistently defined. In contrast, conventional thermodynamics is not subject to such constraints, and Legendre transformations may be freely performed among any thermodynamically conjugate variables to define arbitrary ensembles.

Unlike electromagnetic fields sourced by external matter, pure gravity can also carry conserved charges beyond mass, such as angular momentum. However, due to the highly nonlinear nature of gravity, it is challenging to identify a covariant total derivative term that enables a Legendre transformation between angular momentum and angular velocity. In fact, it remains unclear whether such a transformation would be consistent with the variational principle. However, it is important to analyze the thermodynamic stability of rotating black holes within an ensemble of fixed angular momentum in the context of issues such as the Weak Gravity Conjecture (WGC)\cite{Cheung:2019cwi,Ma:2020xwi,Ma:2022gtm}. This requires a careful examination, within a pure gravity framework, of the validity and consistency of performing a Legendre transformation for angular momentum and angular velocity, a thermodynamic conjugate pair that arises solely from the pure gravitational sector. Due to these uncertainties, it is worthwhile to revisit the thermodynamic stability of several common black objects.

The method of conserved charges provides another effective approach to computing black hole thermodynamics. The Killing vectors of the spacetime metric yield the mass and angular momenta of the black hole, while the \(U(1)\) gauge symmetry of the Maxwell field defines the electric charge.
The covariant phase space formalism developed by Wald \cite{Iyer:1994ys,Wald:1993nt,Wald:1999wa}, known as the Wald formalism, shows that in any covariant gravity theory, each Killing vector $\xi$ defines a corresponding conserved charge. Moreover, these charges satisfy a differential identity, which precisely encodes the first law of black hole mechanics. In contrast to the Euclidean on-shell action method, which relies on a suitable reference background and specific boundary terms, the Wald formalism is largely free of such ambiguities. It is therefore widely employed to define and compute black hole thermodynamic quantities. Notably, the Wald entropy formula \cite{Wald:1993nt} follows directly from this formalism and offers a universally applicable and robust approach. Although the Wald entropy formula has been shown to be a reliable way of reading off the black hole entropy in most black hole examples, the formula turns out to be incorrect for black holes in Horndeski theory \cite{Feng:2015oea,Feng:2015wvb} and Einstein-bumblebee gravity \cite{An:2024fzf,Li:2025tcd}, where fields can be divergent on the horizon. In such situations, one must instead resort to the original Wald formalism, which gives a sympletic first-order differential identity. In this framework, it is crucial to compute other thermodynamic quantities independently, including the electric and magnetic charges, together with their corresponding potentials, as well as the angular momentum and angular velocity in rotating spacetime. Consequently, extracting these quantities reliably from the Wald formalism becomes a central issue. How to obtain the electromagnetic contributions from the covariant phase space has been demonstrated in \cite{Lu:2013ura,Kim:2023ncn}. In this work we focus on extracting the angular momentum and angular velocity of black holes from the perspective of Kaluza-Klein reduction.

While the Wald formalism provides a rigorous framework for computing black hole thermodynamic quantities and ensures that the resulting physical observables satisfy the first law of black hole mechanics, its application becomes nontrivial in low-energy effective theories of string theory and supergravity. This is due to the presence of non-covariant Chern-Simons terms, which complicate the direct implementation of the formalism. In such cases, the Wald formalism must be appropriately modified to account for these non-covariant terms \cite{Tachikawa:2006sz,Barnich:2005kq}. As shown in \cite{Ma:2022nwq}, the non-covariance of the Chern-Simons term introduces a total derivative ambiguity in the otherwise equivalent formulations of the Wald formalism. Although this ambiguity does not affect the formal equivalence, it implies that for actions containing Chern-Simons terms, the lack of a manifest \(U(1)\) gauge invariance renders certain equivalence statements less manifest. In particular, the expression for the electric charge obtained from the Wald formalism does not coincide with the closed-form result derived from the equations of motion. Likewise, the evaluation of the on-shell action depends on the choice of gauge. These issues are analyzed in detail in the following sections.

The structure of this paper is as follows. In Section \ref{The first law and on-shell action in Einstein-Maxwell theory}, we take Einstein-Maxwell theory as an example to discuss the consistency of boundary condition choices within the framework of the variational principle, with particular emphasis on the 4D case where magnetic charge is allowed. We also introduce the definitions of magnetic charge and potential using the Wald formalism. Inspired by the Legendre transformation of the Maxwell field with \(U(1)\) gauge symmetry discussed in Section~\ref{The first law and on-shell action in Einstein-Maxwell theory}, in Section~\ref{Legendre transformation of pure geometry quantities} we introduce total derivative terms for metrics with a \(U(1)\) isometry, in the sense of Kaluza-Klein reduction, to implement a Legendre transformation by adjusting the boundary conditions. The latter part of Section \ref{Legendre transformation of pure geometry quantities}, together with Section \ref{Legendre transformation of the Kerr black hole}, is devoted to testing the effectiveness of our method through several representative examples. We also demonstrate that fixing $(Q_e, Q_m)$ or $(\Phi_e, \Phi_m)$ simultaneously leads to boundary conditions that are incompatible with the variational principle. In Section \ref{FFA}, we explore how Chern-Simons terms affect both the Wald formalism and the Euclidean on-shell action method in the context of five-dimensional minimal supergravity, and propose possible resolutions to the issues that arise. In Section \ref{More about the $U(1)$ gauge choice in Chern-Simons Euclidean action}, we will show that in Chern-Simons theory, it is crucial to choose an appropriate gauge such that the Maxwell field is well-defined at all degenerate points of the spacetime geometry. Only with this condition does the resulting on-shell action correspond to a free energy that is consistent with the first law.
 We conclude in Section \ref{Conclusions} with a summary of our results.

\section{The first law and on-shell action in Einstein-Maxwell theory}\label{The first law and on-shell action in Einstein-Maxwell theory}

The Einstein-Maxwell theory is one of the most commonly studied models of gravity coupled to matter. Despite its simplicity, it captures a wide range of essential features. In four dimensions, the Maxwell field can carry magnetic charge, allowing the RN and Kerr-Newman black holes to be generalized to dyonic black holes.

In this section, using the Einstein-Maxwell theory as a representative example, we analyze black hole thermodynamics through both the Wald formalism and the on-shell action approach. We present explicit definitions of the magnetic potential and magnetic charge that are consistent with the first law of thermodynamics, and discuss how total derivative terms can modify the boundary conditions of the Maxwell sector, thereby changing the thermodynamic ensemble of the black hole.

\subsection{Wald formalism and the first law}

Let us begin with the Einstein-Maxwell theory in arbitrary dimensions
\bea
S_{\mathrm{EM}}=\frac{1}{16\pi}\int d^Dx\sqrt{-g}\mathcal{L}_{\mathrm{EM}}\,,\qquad \mathcal{L}_{\mathrm{EM}}=R-\frac{1}{4}F^2\label{D-dim EM}
\eea
or
\bea
e^{-1}\mathcal{L}_{\mathrm{EM}}=R*\oneone-\frac{1}{2}*F_{\2}\wedge F_{\2}\,.\label{D-dim EM form}
\eea
After performing a variation for this theory, we obtain its equations of motion (EOMs) along with the surface term
\bea
\frac{\delta(\sqrt{-g}\mathcal{L}_{\mathrm{EM}})}{\sqrt{-g}}&=&E_{\mu\nu}\delta g^{\mu\nu}+S_A^\mu\delta A_\mu+\nabla_\mu\Theta_G^\mu+\nabla_\mu\Theta_A^\mu\,,\cr
E_{\mu\nu}&=&R_{\mu\nu}-\frac{1}{2}F^2_{\mu\nu}-\frac{1}{2}g_{\mu\nu}\mathcal{L}_{\mathrm{EM}}\,,\qquad
\Theta_G^\mu=g^{\mu\alpha}\nabla^\beta\delta g_{\alpha\beta}-g^{\alpha\beta}\nabla^\mu\delta g_{\alpha\beta}\,,\cr
\mathbf{S}_A&=&d\big((-1)^{D-2}*F_{\2}\big)\,,\qquad \mathbf{\Theta}_A=(-1)^{D-1}*F_{\2}\wedge\delta A_{\1}\,.\label{EOM and surface EM}
\eea

Within the Wald covariant phase space formalism \cite{Iyer:1994ys,Wald:1993nt}, for any transformation generated by $\xi$, one can define the Noether charge $\mathbf{Q}$ of the theory \eqref{D-dim EM}
\bea
\mathbf{Q}&=&\mathbf{Q}_G+\mathbf{Q}_A\,,\cr
\mathbf{Q}_G&=&-*d\xi\,,\qquad \mathbf{Q}_A=-*F_{\2}\big(i_\xi A_{\1}\big)\,,
\eea
from which the infinitesimal Hamiltonian of the system can be constructed via a codimension-2 hypersurface integral
\bea
\delta \mathcal{H}=\frac{1}{16\pi}\int_{\Sigma^{D-1}}d\big(\delta\mathbf{Q}-i_\xi\mathbf{\Theta}\big)=\frac{1}{16\pi}\oint_{\Sigma^{D-2}}\big(\delta\mathbf{Q}-i_\xi\mathbf{\Theta}\big)\,.
\eea
If $\xi$ is a Killing vector, i.e., $\delta_\xi g_{\mu\nu} = 0$, $\delta_\xi A_{\mu} = 0$, then the infinitesimal Hamiltonian $\delta \mathcal{H}$ vanishes, and the above closed-loop integral can be decomposed into contributions from two codimension-2 hypersurfaces
\bea
\int_{\Sigma^{D-2}_1}\big(\delta\mathbf{Q}-i_\xi\mathbf{\Theta}\big)=\int_{\Sigma^{D-2}_2}\big(\delta\mathbf{Q}-i_\xi\mathbf{\Theta}\big)\label{1st}
\eea
If we place $\Sigma^{D-2}_1$ at asymptotic infinity, the resulting integral can be interpreted as the black hole's mass $M$ and angular momentum $J$
\bea
\int_{\Sigma^{D-2}_1}\big(\delta\mathbf{Q}-i_\xi\mathbf{\Theta}\big)=\delta M-\Omega_H\delta J\,,
\eea
similarly, placing $\Sigma^{D-2}_2$ on the black hole horizon yields the entropy and electric charge\footnote{Here, we choose a gauge in which the electric potential vanishes at asymptotic infinity.}
\bea
\int_{\Sigma^{D-2}_2}\big(\delta\mathbf{Q}-i_\xi\mathbf{\Theta}\big)=T\delta S+\Phi_e\delta Q_e\,.
\eea
Therefore, equation \eqref{1st} provides a derivation of the first law of black hole thermodynamics.

Just as the Wald entropy can be identified from the Wald formalism \cite{Wald:1993nt}, the electric charge $Q_e$ and potential $\Phi_e$ can likewise be extracted from it. In the Maxwell sector, we observe that
\bea
\delta \mathbf{Q}_A-i_\xi\mathbf{\Theta}_A=-\delta*F_{\2}\big(i_\xi A_{\1}\big)+(-1)^Di_\xi*F_{\2}\wedge\delta A_{\1}\,.\label{potential and charge}
\eea
The first term in the above expression exactly corresponds to the definition of charge and potential
\bea
Q_e=\frac{1}{16\pi}\int_{\Sigma^{D-2}}*F_{\2}\,,\qquad \Phi_e=i_\xi A_{\1}|_{r=r_h}^{r\rightarrow\infty}\,,\label{def electric charge and potential}
\eea
while the second term vanishes in pure electric case. From the Maxwell equations \eqref{EOM and surface EM}, we know that $*F_{\2}$ is a closed form, which means that $\Sigma^{D-2}$ can be any hypersurface located anywhere in the spacetime. This is consistent with the principles of electromagnetism, as the charge is independent of the choice of Gaussian surface. Therefore, \eqref{potential and charge} captures the electric part in the first law
\bea
\delta \mathbf{Q}_A-i_\xi\mathbf{\Theta}_A\sim\Phi_e\delta Q_e\,.
\eea

In $D = 4$, black holes can carry magnetic charge, as in the well-known RN-dyonic and Kerr-Newman-dyonic solutions. In this case, for a Killing vector $\xi$, we have $\mathcal{L}_\xi *F_{\2} = 0$, which allows us to define a scalar field $\Psi$
\bea
\mathcal{L}_\xi *F_{\2} =(di_\xi+i_\xi d)*F_{\2} =di_\xi*F_{\2}\,,\quad\Rightarrow\quad i_\xi*F_{\2}=d\Psi\,.\label{magnetic potential}
\eea
Because the Wald formalism \cite{Iyer:1994ys} admits an ambiguity up to an exact form, and following results from previous literature \cite{Lu:2013ura,Ma:2022nwq,Kim:2023ncn}, we may add $d(-\Psi \, \delta A_{\1})$ to equation \eqref{potential and charge} without changing the value of the integral. This yields the following result
\bea
\delta \mathbf{Q}_A-i_\xi\mathbf{\Theta}_A-d(\Psi \, \delta A_{\1})=-\delta*F_{\2}\big(i_\xi A_{\1}\big)-\Psi\delta F_{\2}\,.\label{magnetic potential 2}
\eea
The second term now precisely defines the magnetic charge and potential. Likewise, owing to the Bianchi identity $dF_{\2} = 0$, the magnetic charge is independent of the choice of hypersurface
\bea
Q_m=\frac{1}{16\pi}\int F_{\2}\,,\qquad \Phi_m=\Psi|_{r=r_h}^{r\rightarrow\infty}\,.\label{magnetic potential wald}
\eea
Indeed, this result is fully consistent with the conventional approach to defining the magnetic potential \cite{Rasheed:1997ns,Cano:2020qhy}. From the Maxwell EOM \eqref{EOM and surface EM}, we can define a new 1-form field $\tilde{A}_{\1}$
\bea
d*F_{\2}=0\,,\qquad \Rightarrow\qquad *F_{\2}=d\tilde{A}_{\1}\,.\label{dual A definition}
\eea
Similarly, $\xi$ is also its Killing vector field, satisfying
\bea
\mathcal{L}_\xi\tilde{A}_{\1}=0=(di_\xi+i_\xi d)\tilde{A}_{\1}\,.
\eea
From the definition of the scalar $\Psi$ \eqref{magnetic potential}, we have that
\bea
i_\xi*F_{\2}=i_\xi d\tilde{A}_{\1}=-di_\xi\tilde{A}_{\1}\,,\qquad \Rightarrow\qquad \Psi=-i_\xi\tilde{A}_{\1}\,.
\eea
It can be seen that the Wald formalism (\ref{magnetic potential 2},\ref{magnetic potential wald}) precisely reproduces the expression for the magnetic potential given in the original reference \cite{Rasheed:1997ns,Cano:2020qhy}
\bea
\Phi_m=-i_\xi\tilde{A}_{\1}|_{r=r_h}^{r\rightarrow\infty}\,.\label{magnetic potential wald 2}
\eea
In other words, the Wald formalism offers precise and rigorous definitions of the electric and magnetic charges $(Q_e, Q_m)$, along with their corresponding potentials $(\Phi_e, \Phi_m)$, in a way that is fully consistent with the first law (\ref{magnetic potential},\ref{magnetic potential 2}).

It is important to emphasize that the manifest consistency between the electric charge definition given by the Wald formalism \eqref{potential and charge} and that derived from the EOMs \eqref{EOM and surface EM} is a direct consequence of the manifest gauge invariance of the action. In cases like the Chern-Simons (CS) action, where the EOMs are $U(1)$ gauge invariant, but the action shifts by a total derivative under this gauge transformation. In this case, the Maxwell sector in the Wald formalism no longer provides a closed form. Consequently, a well-defined expression for the charge cannot be obtained. How to use Wald formalism to define charge in Chern-Simons theory will be addressed in detail in Section \ref{FFA}.

\subsection{Euclidean on-shell action and different boundary condition}

Let us now shift our focus to the Euclidean on-shell action. Gibbons and Hawking observed that \cite{Gibbons:1976ue}, to obtain a finite and physically meaningful free energy, one must introduce appropriate boundary terms in the gravitational action to impose well-defined boundary conditions, along with counterterms to eliminate divergences. This led to the formulation of the Gibbons-Hawking-York boundary term.

Unlike the purely gravitational sector, which requires careful treatment, the on-shell action of the Maxwell sector is already finite, and the choice of appropriate total derivative terms is equivalent to performing a Legendre transformation. Legendre transformation allows the same black hole solution to be examined in different thermodynamic ensembles, with the resulting on-shell action naturally yielding different types of free energy.

Similarly, taking the Einstein-Maxwell theory in arbitrary dimensions as an example, we may add a total derivative term
\bea
I_\gamma=-\frac{1}{16\pi}\int d\big(A_{\1}\wedge*F_{\2}\big)\label{total derivative EM}
\eea
to the action\footnote{Since we are considering the Euclidean action, an extra minus sign appears compared to the earlier expression in \eqref{D-dim EM form}.}
\bea
I_{A}=-\frac{1}{16\pi}\int \Big[-\frac{1}{2}*F_{\2}\wedge F_{\2}\Big]
\eea
which does not affect the EOMs and hence leaves the black hole solution unmodified. We perform a variation of $I_A + \gamma I_\gamma$ and simplify the result using the EOMs, obtaining
\bea
\delta(I_A + \gamma I_\gamma)&=&-\frac{\beta}{16\pi}\int_{\Sigma^{D-1}}\big[
(\gamma-1)\delta A_{\1}\wedge*F_{\2}+\gamma A_{\1}\wedge\delta *F_{\2}
\big]\,.\label{variation}
\eea
Here, $\beta$ denotes the Euclidean time period, which is the inverse of the temperature, $\beta = T^{-1}$. Let us now consider the RN black hole in arbitrary dimensions as an example. Upon substituting the definitions of the electric charge and electric potential given in ~\eqref{def electric charge and potential}, this part of the integral is found to evaluate to
\bea
\delta(I_A + \gamma I_\gamma)&=&\beta\big[(\gamma-1)Q_e\delta \Phi_e+\gamma\Phi_e\delta Q_e\big]\,.\label{variation impose}
\eea
This result indicates that \(I_A + \gamma I_\gamma\) is a function of \(\Phi_e\) and \(Q_e\).

 For $\gamma = 0$, where no boundary term is added for the Maxwell field, we are effectively imposing the boundary condition $\delta *F_{\2}|_\partial = 0$, corresponding to a Dirichlet condition. In this case, the on-shell action yields the Gibbs free energy,
\bea
F_\mathrm{G}=\frac{I(T,\Phi_e)}{\beta}=M-TS-\Phi_eQ_e\,,
\eea
which is a thermodynamic potential expressed in terms of the electric potential. Conversely, for $\gamma = 1$, the boundary condition becomes $\delta A_{\1}|_\partial = 0$, corresponding to a Neumann condition. The resulting on-shell action gives the Helmholtz free energy
\bea
F_\mathrm{H}=\frac{I(T,Q_e)}{\beta}=M-TS\,,
\eea
which is a function of the electric charge. Therefore, the total derivative term \eqref{total derivative EM} plays a role of a Legendre transformation multiplier on the thermodynamic system. Here, $\gamma$ can only be 0 or 1; any other value would produce two boundary terms with distinct behaviors upon variation, violating the variational principle \footnote{Of course, exceptions exist. For example, in the context of holographic complexity, it is sometimes desirable to set $\gamma = \frac{1}{2}$ in order to compute the electromagnetic dual on-shell action in Lorentz signature \cite{Goto:2018iay,Liu:2019smx}. This choice is specific to the goals of holographic complexity and does not pertain to the thermodynamic Legendre transformations which require either $\gamma=0$ or $\gamma=1$.}.

As in the previous section, $D = 4$ black holes can carry magnetic charge, and we once again introduce the 1-form field $\tilde{A}_{\1}$ \eqref{dual A definition}. In this case, the variation \eqref{variation} can be rewritten as follows
\bea
\delta(I_A + \gamma I_\gamma)&=&-\frac{\beta}{16\pi}\int_{\Sigma^{3}}\big[
(\gamma-1)\delta A_{\1}\wedge*F_{\2}+\gamma F_{\2}\wedge\delta \tilde{A}_{\1}
\big]\,,\cr
&=&\beta\big[(\gamma-1)Q_e\delta \Phi_e-\gamma Q_m\delta \Phi_m\big]\,.\label{variation impose EMD}
\eea
Likewise, we substitute the expressions for the magnetic charge ~\eqref{magnetic potential wald} and the magnetic potential ~\eqref{magnetic potential wald 2}. As follows from the above analysis of the purely electric case, \(I_A + \gamma I_\gamma\) cannot simultaneously depend on both the electric potential \(\Phi_e\) and the magnetic potential \(\Phi_m\). Equivalently, for 4D RN-dyonic black hole, the on-shell action can only be formulated as a function of either the electric potential and magnetic charge, or the magnetic potential and electric charge
\bea
F_{\gamma=0}&=&F(T,\Phi_e,Q_m)=M-TS-\Phi_e Q_e\,,\cr
F_{\gamma=1}&=&F(T,Q_e,\Phi_m)=M-TS-\Phi_m Q_m\,.\label{Einstein-Maxwell dyonic F}
\eea
Thus, the Euclidean on-shell action for the dyonic black hole free energy does not have the electromagnetic duality regardless the Legendre transformation. This result carries important implications. It shows that, within the framework of the variational principle, the RN-dyonic black hole cannot be consistently treated in an ensemble where both the electric and magnetic charges are fixed, nor in one where both the electric and magnetic potentials are fixed. Therefore, when examining the thermodynamic stability of dyonic black objects, particular attention must be paid to the choice of ensemble. In Section \ref{Legendre transformation of pure geometry quantities}, we shall illustrate this point through an examination of pure electric, pure magnetic and dyonic black hole solutions.

\section{4D EMD Legendre transformation and its 5D Einstein origin}\label{Legendre transformation of pure geometry quantities}

Unlike electric charge, black holes can carry angular momentum even in pure gravity theory. In higher-dimensional spacetimes, it is also common to encounter solutions such as boosted black strings, which carry linear momentum. These physical quantities also participate in the thermodynamic process, with the free energy derived from the on-shell action depending on the angular or linear velocity.

In linear Maxwell theory, it is relatively straightforward to construct a covariant total derivative term that implements a Legendre transformation between the thermodynamic conjugate pair of electric (or magnetic) charge and potential \eqref{total derivative EM}. In contrast, due to the highly nonlinear nature of gravity, it is much more challenging to identify an explicit covariant total derivative term, analogous to equation \eqref{total derivative EM}, that serves as the Legendre transformation term for conjugate pairs such as angular velocity and angular momentum, or linear velocity and linear momentum.

However, for black objects carrying angular or linear momentum, there always exists a Killing vector $\partial_z$, allowing the metric to be written in Kaluza-Klein dimensional reduction form
\be
d\hat{s}_{D+1}^2 = ds_{D}^2 + \Omega (dz + {\cal A}_{\1})^2\,.\label{rotating ansatz}
\ee
From the lower-dimensional viewpoint, the Kaluza-Klein vector ${\cal A}_{\1}$ manifests as a $U(1)$ gauge field, with the higher-dimensional angular or linear momentum reinterpreted as an electric charge. Therefore, in lower dimensions, the electric charge and potential associated with the Kaluza-Klein vector field ${\cal A}_{\1}$ correspond to the linear or angular momentum and the linear or angular velocity of the higher-dimensional black hole. Moreover, if ${\cal A}_{\1}$ exhibits a Dirac-string structure, giving rise to magnetic charge and potential, this signals that the higher-dimensional black hole is a Kaluza-Klein monopole carrying a Misner string. Consequently, as in the Einstein-Maxwell theory, one can perform a Legendre transformation in the reduced theory to transition between different thermodynamic ensembles of the black hole.

Therefore, within the framework of dimensional reduction, a Legendre transformation can be carried out for the angular velocity and momentum conjugate pair. In this section, we demonstrate that for metrics admitting a \(U(1)\) fiber bundle structure of the form \eqref{rotating ansatz}, the total derivative term required for the Legendre transformation takes the form
\bea
\hat{\mathcal{L}}_{\mathrm{Legendre}}=d\left(
\Omega\mathcal{A}_{\1}\wedge\hat{*}\mathcal{F}_{\2}
\right)\,,\qquad \mathcal{F}_{\2}=d\mathcal{A}_{\1}\,  \label{Legendre transformation D}
\eea
and we will demonstrate its validity within the Kaluza-Klein reduction framework. And then, we will present several examples to verify our conclusions.

\subsection{Legendre transformation in $D$-dimensions and $(D+1)$-dimensions}

In this section, we denote fields in the $(D+1)$-dimensional theory with a hat. Starting from $(D+1)$-dimensional general relativity
\bea
\hat{S}=\frac{1}{16\pi }\int d^{D+1}x\sqrt{-\hat{g}} \hat{R}\,,\label{D+1}
\eea
we perform a circle reduction along the Killing vector
\bea
d\hat{s}_{D+1}^2=e^{-\sqrt{\frac{2}{(D-1)(D-2)}}\phi}ds_D^2+e^{\sqrt{\frac{2(D-2)}{D-1}}\phi}(dz+\mathcal{A}_{\1})^2\,,\label{reduction ansatz D}
\eea
resulting in a $D$-dimensional Einstein-Maxwell-dilaton (EMD) theory \footnote{A careful examination of our subsequent analysis reveals that fixing the low-dimensional theory in the Einstein frame is not fundamentally required, though we employ this convention in the present discussion for computational simplicity in lower dimensions.}
\bea
S_{\mathrm{EMD}}&=&\frac{1}{16\pi }\int d^Dx\sqrt{-g} \mathcal{L}_{\mathrm{EMD}}\,,\qquad
\mathcal{L}_{\mathrm{EMD}}=R-\frac{1}{2}(\partial\phi)^2-\frac{1}{4}e^{\sqrt{\frac{2(D-1)}{D-2}}\phi}\mathcal{F}^2\label{EMD D}
\eea
or in form
\bea
e^{-1}\mathcal{L}_{\mathrm{EMD}}&=&R* 1-\frac{1}{2}*d\phi\wedge d\phi-\frac{1}{2}e^{\sqrt{\frac{2(D-1)}{D-2}}\phi}*\mathcal{F}_{\2}\wedge \mathcal{F}_{\2}\,.
\eea

The $U(1)$ fiber bundle present in the metric reduces to a $U(1)$ gauge field, which is a Maxwell field in the lower-dimensional theory. Consequently, we can employ the method introduced in previous Section \ref{The first law and on-shell action in Einstein-Maxwell theory} to calculate the electric potential and charge of the $\mathcal{A}_\mu$ field, and use the variational principle
\bea
\delta\big(e^{-1}\mathcal{L}_{\mathrm{EMD}}\big)&\sim&-e^{\sqrt{\frac{2(D-1)}{D-2}}\phi}*\mathcal{F}_{\2}\wedge \delta\mathcal{F}_{\2}= d\Big(-e^{\sqrt{\frac{2(D-1)}{D-2}}\phi}\delta\mathcal{A}_{\1}\wedge *\mathcal{F}_{\2}\Big)\,
\eea
to identify the appropriate total derivative term required
\bea
e^{-1}\mathcal{L}_{\mathrm{Legendre}}=d\Big(e^{\sqrt{\frac{2(D-1)}{D-2}}\phi} \mathcal{A}_{\1}\wedge*\mathcal{F}_{\2}\Big)\,.\label{Legendre EMD}
\eea
for the Legendre transformation. In fact, the electric charge and potential associated with $\mathcal{A}_\mu$ correspond to the linear or angular momentum and the linear or angular velocity, respectively, in the higher-dimensional theory \eqref{D+1}. The total derivative term \eqref{Legendre EMD} therefore serves, from the lower-dimensional viewpoint, as the Legendre transformation expression for these geometrically induced physical quantities.

It is precisely the well-founded justification for performing the Legendre transformation in the Maxwell sector of $D$-dimensional EMD theory that gives us confidence in extending this procedure to $(D+1)$-dimensional pure gravity. Although writing down a covariant expression for the total derivative term is generally challenging, if the $(D+1)$-dimensional metric admits a $U(1)$ fiber bundle structure (\ref{rotating ansatz},\ref{reduction ansatz D}), Kaluza-Klein reduction ensures the existence of a corresponding Legendre transformation, as expressed in \eqref{Legendre EMD}. We now verify that the $(D+1)$-dimensional counterpart of \eqref{Legendre EMD} is given by \eqref{Legendre transformation D}.

 According to the general formula \eqref{Legendre transformation D}, the Legendre transformation in $(D+1)$-dimensional general relativity takes the form
 \bea
\hat{\mathcal{L}}_{\mathrm{Legendre}}=\hat{\nabla}_{\mu}\Big(e^{\sqrt{\frac{2(D-2)}{D-1}}\phi}
\mathcal{F}^{\mu\nu}\mathcal{A}_{\nu}
\Big)\,,\quad \Leftrightarrow\quad  \hat{e}^{-1}\hat{\mathcal{L}}_{\mathrm{Legendre}}=d\Big(
e^{\sqrt{\frac{2(D-2)}{D-1}}\phi}\mathcal{A}_{\1}\wedge\hat{*}\mathcal{F}_{\2}
\Big)\,.\label{Legendre GR}
\eea
We will now employ dimensional reduction to demonstrate the equivalence between \eqref{Legendre EMD} and \eqref{Legendre GR}. Because of the reduction ansatz \eqref{reduction ansatz D}, the natural choice of vielbein basis $\hat{e}^A=\{\hat{e}^a,\hat{e}^z\}$ are
\bea
\hat{e}^a=e^{-\frac{\phi}{\sqrt{2(D-1)(D-2)}}}e^a\,,\qquad \hat{e}^z=e^{\sqrt{\frac{D-2}{2(D-1)}}\phi}\big(dz+\mathcal{A}_{\1}\big)\,.
\eea
The 2-form field strength is
\bea
\mathcal{F}_{\2}=\frac{1}{2}\mathcal{F}_{ab}e^a\wedge e^b
=\frac{1}{2}e^{\sqrt{\frac{2}{(D-1)(D-2)}}\phi}\mathcal{F}_{ab}\hat{e}^a\wedge \hat{e}^b
\eea
and its Hodge dual is
\bea
\hat{*}\mathcal{F}_{\2}&=&\frac{1}{2}e^{\sqrt{\frac{2}{(D-1)(D-2)}}\phi}\mathcal{F}_{ab}\,\hat{*}\,(\hat{e}^a\wedge \hat{e}^b)\cr
&=&\frac{1}{2}e^{\sqrt{\frac{2}{(D-1)(D-2)}}\phi}\mathcal{F}^{ab}\frac{1}{(D-1)!}\epsilon_{C_1\cdots C_{D-2}C_{D-1}ab}
\hat{e}^{C_1}\wedge\cdots\wedge\hat{e}^{C_{D-2}}\wedge \hat{e}^{C_{D-1}}\cr
&=&\frac{1}{2}e^{\sqrt{\frac{2}{(D-1)(D-2)}}\phi}\mathcal{F}^{ab}\frac{1}{(D-2)!}\epsilon_{C_1\cdots C_{D-2}zab}
\hat{e}^{C_1}\wedge\cdots\wedge\hat{e}^{C_{D-2}}\wedge \hat{e}^{z}\cr
&=&\frac{1}{2}e^{\sqrt{\frac{2}{(D-1)(D-2)}}\phi}\mathcal{F}^{ab}\frac{1}{(D-2)!}\epsilon_{C_1\cdots C_{D-2}ab}
e^{C_1}\wedge\cdots\wedge e^{C_{D-2}}\wedge \big(dz+\mathcal{A}_{\1}\big)\cr
&=&e^{\sqrt{\frac{2}{(D-1)(D-2)}}\phi}*\mathcal{F}_{\2}\wedge \big(dz+\mathcal{A}_{\1}\big)\,.
\eea

The $(D+1)$-dimensional Legendre total derivative \eqref{Legendre GR} reduces to
\bea
\hat{e}^{-1}\hat{\mathcal{L}}_{\mathrm{Legendre}}&=&d\Big(
e^{\sqrt{\frac{2(D-2)}{D-1}}\phi}\mathcal{A}_{\1}\wedge\,\hat{*}\,\mathcal{F}_{\2}
\Big)\cr
&=&d\Big(
e^{\sqrt{\frac{2(D-1)}{D-2}}\phi}\mathcal{A}_{\1}\wedge*\mathcal{F}_{\2}\wedge \big(dz+\mathcal{A}_{\1}\big)
\Big)\cr
&=&d\Big(
e^{\sqrt{\frac{2(D-1)}{D-2}}\phi}\mathcal{A}_{\1}\wedge*\mathcal{F}_{\2}
\Big)\wedge dz\cr
&=&e^{-1}\mathcal{L}_{\mathrm{Legendre}}\wedge dz\,.
\eea
Hence, \eqref{Legendre EMD} and \eqref{Legendre GR} are equivalent within the framework of Kaluza-Klein reduction. In the rest of this section, we will verify the correctness of this expression \eqref{Legendre transformation D} through explicit examples.

\subsection{Pure electric/boost and pure magnetic/monopole in 4D/5D}

The 4D EMD theory
\bea
S_4&=&\frac{1}{16\pi}\int d^4x\sqrt{-g} \mathcal{L}_{\mathrm{EMD}}\,,\qquad
\mathcal{L}_{\mathrm{EMD}}=R-\frac{1}{2}(\partial\phi)^2-\frac{1}{4}e^{\sqrt{3}\phi}F^2\,.\label{EMD}
\eea
admits a pair of static black hole solutions with identical metric forms
\bea
ds_4^2&=&-H^{-\frac{1}{2}}fdt^2+H^{\frac{1}{2}}\left(f^{-1}dr^2+r^2d\Omega_2^2\right)\,,\cr
H&=&1+\frac{q}{r},\quad f=1-\frac{\mu}{r}\,.
\eea
While the specific forms of the Maxwell and dilaton fields determine whether the solution describes a pure electric or pure magnetic black hole
\bea
\text{Pure electric}:\,  A_{\1}&=&\omega dt\,,\quad \phi=\frac{\sqrt{3}}{2}\log H\,,\quad  \omega=\frac{\sqrt{q(\mu+q)}}{rH}\,,\cr
\text{Pure magnetic}:\,  A_{\1}&=&\sqrt{q(\mu+q)}\cos\theta d\varphi\,,\quad \phi=-\frac{\sqrt{3}}{2}\log H\,.\label{Electric and Magnetic}
\eea

In \cite{Liu:2022wku}, we conducted a thorough analysis of the thermodynamics of both these two solutions using both the Wald formalism and the on-shell action approach. From the Noether charge and surface term of the Maxwell sector,
\bea
\mathbf{Q}_A=-e^{\sqrt{3}\phi}*F_{\2}\big(i_\xi A_{\1}\big)\,,\qquad \mathbf{\Theta}_A=-e^{\sqrt{3}\phi}*F_{\2}\wedge\delta A_{\1}\,,
\eea
the Wald formalism
\bea
\delta \mathbf{Q}_A-i_\xi\mathbf{\Theta}_A=-\delta\big(e^{\sqrt{3}\phi}*F_{\2}\big)\big(i_\xi A_{\1}\big)
+i_\xi\big(e^{\sqrt{3}\phi}*F_{\2}\big)\wedge\delta A_{\1}\label{potential and charge EMD}
\eea
yields precise definitions of the electric charge and potential. The result is in full agreement with that derived from the Maxwell equations $\mathbf{S}_A=d\big(e^{\sqrt{3}\phi}*F_{\2}\big)$. Similarly, guided by the EOMs, we introduce the dual 1-form field $\tilde{A}_{\1}$ and the scalar field $\Psi$
\bea
e^{\sqrt{3}\phi}*F_{\2}=d\tilde{A}_{\1}\,,\qquad \Psi=-i_\xi \tilde{A}_{\1}\,.\label{1-form EMD}
\eea
By utilizing the ambiguity in the Wald formalism,
\bea
\delta \mathbf{Q}_A-i_\xi\mathbf{\Theta}_A+d(\Psi\delta A_{\1})=-\delta\big(e^{\sqrt{3}\phi}*F_{\2}\big)\big(i_\xi A_{\1}\big)-\Psi\delta F_{\2}\,,
\eea
we can derive precise definitions of the magnetic charge and magnetic potential, which are fully consistent with the results obtained from the Bianchi identity $dF_{\2}=0$.

We find that the thermodynamics of this pair of solutions are fully equivalent
\bea
M&=&\frac{\Omega_2}{16\pi }(q+2\mu)\,,\quad T=\frac{1}{4\pi}\frac{1}{\sqrt{\mu(\mu+q)}}\,,\quad S=\frac{\Omega_2}{4}\mu^{\frac{3}{2}}\sqrt{\mu+q}\,,\cr
\Phi_{e,m}&=&\sqrt{\frac{q}{\mu+q}}\,,\quad Q_{e,m}=\frac{\Omega_2}{16\pi }\sqrt{q(\mu+q)}\,,
\eea
rendering it impossible to distinguish between pure electric and pure magnetic black holes based only on thermodynamic quantities and the first law. However, after computing the on-shell action,
\bea
\text{Pure\ electric}:\, F_\mathrm{H}&=&F(T,\Phi_e)=M-TS-\Phi_e Q_e\,,\cr
\text{Pure\ magnetic}:\, F_\mathrm{G}&=&F(T,Q_m)=M-TS\,,
\eea
we find that the two black holes reside in distinct thermodynamic ensembles.

\subsubsection{Legendre transformation in 4D EMD theory}
When studying the thermodynamic stability of black holes, it is crucial to specify a particular ensemble. From the viewpoint of the action, this amounts to altering the boundary conditions through a Legendre transformation.

Similar to the Einstein-Maxwell case \eqref{total derivative EM}, in the EMD theory \eqref{EMD}, we can introduce the following total derivative term
\bea
e^{-1}\mathcal{L}_{\mathrm{Legendre}}=d\big(e^{\sqrt{3}\phi}*F_{\2}\wedge A_{\1}\big)\,,\qquad \Leftrightarrow\qquad
\mathcal{L}_{\mathrm{Legendre}}=\nabla_\mu\big(e^{\sqrt{3}\phi}F^{\mu\nu}A_\nu\big)\,.\label{Legendre}
\eea
to change the boundary conditions
\bea
\delta\big(e^{-1}( \mathcal{L}_{\mathrm{EMD}}+\gamma \mathcal{L}_{\mathrm{Legendre}})\big)&=&
d\Big(\gamma\delta(e^{\sqrt{3}\phi}*F_{\2})\wedge A_{\1}+(\gamma-1)e^{\sqrt{3}\phi}*F_{\2}\wedge\delta A_{\1}\Big)\,.\label{Legendre transformation}
\eea
It can be observed that different values of $\gamma$ correspond to different boundary conditions
\bea
&&\gamma=0,\quad \Rightarrow\quad \delta(e^{\sqrt{3}\phi}*F_{\2})=0,\quad \Rightarrow\quad  \mathrm{Dirichlet\ condition},\cr
&&\gamma=1,\quad \Rightarrow\quad \delta A_{\1}=0,\quad \Rightarrow\quad  \mathrm{Neumann\ condition}.
\eea
In Euclidean version, the different boundary conditions correspond to different thermodyanmic function. For pure electric black holes, when we impose the Dirichlet boundary conditions, the resulting Euclidean on-shell action is a functional of temperature $T$ and electric potential $\Phi_e$; under Neumann boundary conditions, it becomes a functional of temperature $T$ and electric charge $Q_e$
\bea
F_{\gamma=0}&=&F(T,\Phi_e)=M-TS-\Phi_e Q_e\,,\cr
F_{\gamma=1}&=&F(T,Q_e)=M-TS\,.
\eea

The discussion for pure magnetic black holes is similar to that for pure electric ones. However, following the example of 4D Einstein-Maxwell theory \eqref{dual A definition}, we can introduce the dual 1-form field $\tilde{A}_{\1}$ via \eqref{1-form EMD}. Therefore, the variation \eqref{Legendre transformation} can be rewritten as follows\footnote{Since we are considering pure magnetic black holes rather than dyonic ones, we have rewritten the $(\gamma - 1)$ term using the dual 1-form field, in contrast to the expression in \eqref{dual A definition}.}
\bea
\delta\big(e^{-1}( \mathcal{L}_{\mathrm{EMD}}+\gamma \mathcal{L}_{\mathrm{Legendre}})\big)&=&
d\Big(\gamma\delta\tilde{A}_{\1}\wedge F_{\2}+(\gamma-1)\tilde{A}_{\1}\wedge\delta F_{\2}\Big)\,.
\eea
From this reformulated expression, it is clear that the situation for pure magnetic black holes is the exact opposite of the pure electric case: with Dirichlet boundary conditions, the on-shell action is a functional of temperature and magnetic charge; while with Neumann boundary conditions, it becomes a functional of temperature and magnetic potential
\bea
F_{\gamma=0}&=&F(T,Q_m)=M-TS\,,\cr
F_{\gamma=1}&=&F(T,\Phi_m)=M-TS-\Phi_m Q_m\,.
\eea

\subsubsection{Lifting to 5D}

The 4D EMD theory \eqref{EMD} can be uplifted to 5D general relativity
\bea
\hat{S}_5=\frac{1}{16\pi }\int d^5x\sqrt{-\hat{g}} \hat{R}\label{GR 5D}
\eea
via the following reduction ansatz
\bea
d\hat{s}_5^2=e^{-\frac{1}{\sqrt{3}}\phi}ds_4^2+e^{\frac{2}{\sqrt{3}}\phi}\left(dz+A_{\1}\right)^2\,.\label{reduction ansatz}
\eea

The pure electric charged black hole corresponds to the boost black string
\bea
d\hat{s}_5^2&=&-H^{-1}fdt^2+\frac{dr^2}{f}+r^2d\Omega_2^2+H\left(dz+\omega dt\right)^2\,,\label{boost}
\eea
while the pure magnetic black hole corresponds to the Kaluza-Klein monopole \eqref{Electric and Magnetic}
\bea
d\hat{s}_5^2&=&-fdt^2+H\Big(\frac{dr^2}{f}+r^2d\Omega_2^2
\Big)+H^{-1}\Big(dz+\sqrt{q(\mu+q)}\cos\theta d\varphi\Big)^2\,.\label{monopole}
\eea

As in the four-dimensional case, the boosted black string and the Kaluza-Klein monopole in five dimensions share the same thermodynamic conjugate pairs: the boosted string's angular momentum and angular velocity $(Q_v,\Phi_v)$ correspond directly to the monopole's NUT charge and NUT potential $(Q_p,\Phi_p)$
\bea
M&=&\frac{\Omega_2}{16\pi }(q+2\mu)\,,\quad T=\frac{1}{4\pi}\frac{1}{\sqrt{\mu(\mu+q)}}\,,\quad S=\frac{\Omega_2}{4}\mu^{\frac{3}{2}}\sqrt{\mu+q}\,,\cr
\Phi_{v,p}&=&\sqrt{\frac{q}{\mu+q}}\,,\quad Q_{v,p}=\frac{\Omega_2}{16\pi }\sqrt{q(\mu+q)}\,.\label{boost and KK}
\eea
Consequently, these two black objects are indistinguishable from the view of black hole thermodynamics and the first law. Similarly, the on-shell action indicates that these two five-dimensional black objects belong to distinct thermodynamic ensembles
\bea
\text{Boost\ black string}:\, F_\mathrm{H}&=&F(T,\Phi_v)=M-TS-\Phi_v Q_v\,,\cr
\text{Kaluza-Klein\ monopole}:\, F_\mathrm{}&=&F(T,Q_p)=M-TS\,.
\eea

Based on our proposed method \eqref{Legendre transformation D} and the reduction ansatz \eqref{reduction ansatz}, we can perform a Legendre transformation for black objects in 5D pure general relativity \eqref{GR 5D} by introducing the following total derivative term
\bea
\hat{e}^{-1}\hat{\mathcal{L}}_{\mathrm{Legendre}}=d\big(e^{\frac{2}{\sqrt{3}}\phi} A_{\1}\wedge\hat{*}\,F_{\2}\big)\,,\qquad\Leftrightarrow\qquad
\hat{\mathcal{L}}_{\mathrm{Legendre}}=\hat{\nabla}_\mu\big(e^{\frac{2}{\sqrt{3}}\phi}F^{\mu\nu}A_\nu\big).\label{Legendre 2}
\eea
It is worth noting that the dilaton field of the boosted black string differs from that of the Kaluza-Klein monopole by an overall sign \eqref{Electric and Magnetic}. With the addition of this term, we successfully change the ensemble, making the on-shell action a functional of the linear momentum in boost case and of the NUT potential in monopole case
\bea
\text{Boost\ black string}:\, F_{\gamma=1}&=&F(T,Q_v)=M-TS\,,\cr
\text{Kaluza-Klein\ monopole}:\, F_{\gamma=1}&=&F(T,\Phi_p)=M-TS-\Phi_p Q_p\,.
\eea

Therefore, we have successfully carried out a Legendre transformation for pure geometric physical quantities \eqref{Legendre transformation D}, thereby confirming the validity and effectiveness of our proposed method.

\subsection{4D dyonic black hole}

In this section, we explore the Legendre transformation of 4D dyonic black holes and the appropriate choice of thermodynamic ensemble.

EMD theory admits dyonic black hole solutions \cite{Lu:2013uia} and can be extended to the AdS case \cite{Lu:2013ura} as well as to rotating case \cite{Rasheed:1995zv,Larsen:1999pp}. Here, we focus only on the static case in asymptotically flat spacetime
\bea
ds_4^2&=&-\frac{\Delta_r}{\sqrt{H_1H_2}}dt^2+\sqrt{H_1H_2}\Big(\frac{dr^2}{\Delta_r}+d\theta^2+\sin^2\theta d\varphi^2
\Big)\,,\cr
A_{\1}&=&-\frac{Q(p+2r-\mu)}{H_2}dt-2P\cos\theta d\varphi\,,\qquad \phi=\frac{\sqrt{3}}{2}\log\frac{H_2}{H_1}\,.\label{EMD dyonic}
\eea
Here, the explicit expressions for the relevant functions are given by
\bea
H_1&=&r^2+r (p-\mu )+\frac{p (p-\mu ) (q-\mu )}{2 (p+q)}\,,\qquad
H_2=r^2+r (q-\mu )+\frac{q (p-\mu ) (q-\mu )}{2 (p+q)}\,,\cr
\Delta_r&=&r^2-\mu  r\,,\qquad Q^2=\frac{q \big(q^2-\mu ^2\big)}{4 (p+q)}\,,\qquad P^2=\frac{p \big(p^2-\mu ^2\big)}{4 (p+q)}\,.
\eea
The black hole horizon is located at $r = r_h$, which is defined by the condition $\Delta_r(r_h) = 0$, implying $r_h = \mu$. Here, we highlight the method for computing the magnetic charge and potential. The magnetic charge $Q_m$ is determined by the Bianchi identity
\bea
Q_m=\frac{1}{16\pi}\int_{S^2}F_{\2}=\frac{P}{2}\,.\label{EMD dyonic mag1}
\eea
The dual one-form field $\tilde{A}_{\1}$ \eqref{1-form EMD} takes the form
\bea
\tilde{A}_{\1}=\frac{P(q+2r-\mu)}{H_1}dt-2Q\cos\theta d\varphi\,.
\eea
Then the magnetic potential is
\bea
\Phi_m=-i_\xi\tilde{A}_{\1}|_{r=r_h}^{r\rightarrow\infty}=\frac{2 P (p+q)}{p (\mu +p)}\,.\label{EMD dyonic mag2}
\eea
The other thermodynamic quantities are given by
\bea
M&=&\frac{p+q}{4}\,,\qquad T=\frac{\mu  (p+q)}{2 \pi  \sqrt{pq} (\mu +p) (\mu +q)}\,,\qquad S=\frac{\pi  \sqrt{pq} (\mu +p) (\mu +q)}{2 (p+q)}\,,\cr
\Phi_e&=&\frac{2 Q (p+q)}{q (\mu +q)}\,,\qquad Q_e=\frac{Q}{2}\,,\label{EMD dyonic ele}
\eea
which satisfy the first law
\bea
\delta M=T\delta S+\Phi_e\delta Q_e+\Phi_m\delta Q_m\,.
\eea
The on-shell action can be expressed as a functional of the electric potential and the magnetic charge
\bea
F=F(T,\Phi_e,Q_m)=M-TS-\Phi_eQ_e\,.
\eea

We now proceed to analyze the boundary conditions and perform the Legendre transformation. As in the pure electric or pure magnetic cases, we include the total derivative term \eqref{Legendre} and
\bea
\delta\big(e^{-1}( \mathcal{L}_{\mathrm{EMD}}+\gamma \mathcal{L}_{\mathrm{Legendre}})\big)&=&
d\Big(\gamma\delta(e^{\sqrt{3}\phi}*F_{\2})\wedge A_{\1}+(\gamma-1)\tilde{A}_{\1}\wedge\delta F_{\2}\Big)\,.
\eea

This expression shows that the variational principle does not allow for an ensemble treating both the electric charge $Q_e$ and the magnetic charge $Q_m$ as independent variables. Our calculation reveals that the contribution from the added total derivative term is
\bea
\gamma I_{\mathrm{Legendre}}&=&-\frac{\gamma}{16\pi}\int e^{-1}\mathcal{L}_{\mathrm{Legendre}}=\gamma\beta\big(\Phi_e Q_e-\Phi_m Q_m\big)\,.
\eea
This means that dyonic black holes can only reside in ensembles with either $(\Phi_e, Q_m)$ or $(\Phi_m, Q_e)$ held fixed. The variational principle forbids fixing both charges $(Q_e, Q_m)$ or both potentials $(\Phi_e, \Phi_m)$ simultaneously.

The same conclusion holds after uplifting to 5D Einstein theory \eqref{GR 5D}. A 5D Ricci-flat solution endowed with both a boost and a Kaluza-Klein monopole allows one to uplift the dyonic black hole in 4D EMD theory \eqref{EMD dyonic} to 5D through the reduction ansatz \eqref{reduction ansatz},
\bea
d\hat{s}_5^2&=&-\frac{\Delta_r}{H_2}dt^2+H_1\Big(\frac{dr^2}{\Delta_r}+d\theta^2+\sin^2\theta d\varphi^2
\Big)\cr
&&+\frac{H_2}{H_1}\big(dz-\frac{Q(p+2r-\mu)}{H_2}dt-2P\cos\theta d\varphi\big)^2\,.
\eea
The thermodynamic quantities \((Q_v,\Phi_v)\) arising from the boost sector and \((Q_p,\Phi_p)\) associated with the Kaluza-Klein monopole sector correspond, respectively, to the electric \eqref{EMD dyonic ele} and magnetic (\ref{EMD dyonic mag1},\ref{EMD dyonic mag2}) charges and potentials of the 4D dyonic black hole. Moreover, the Legendre total derivative term \eqref{Legendre 2}
\bea
\gamma\beta^{-1} \hat{I}_{\mathrm{Legendre}}&=&\gamma\big(\Phi_v Q_v-\Phi_p Q_p\big)\,
\eea
further implies that the free energy of this solution can only be a functional of either $(\Phi_v, Q_p)$ or $(\Phi_p, Q_v)$. This is closely analogous to the 4D dyonic case \eqref{Einstein-Maxwell dyonic F}, where the electric and magnetic sectors are not symmetric in the sense of the Euclidean on-shell action. In 5D pure gravity, although the boosted sector and the Kaluza-Klein monopole share identical black hole thermodynamic structures \eqref{boost and KK}, they are nevertheless inequivalent in the black hole free energy defined via the Euclidean on-shell action. As a consequence, a black hole ensemble consistent with the variational principle cannot simultaneously fix $(Q_v, Q_p)$ or $(\Phi_v, \Phi_p)$.


\section{Legendre transformation of the Kerr black hole}\label{Legendre transformation of the Kerr black hole}

Although rotating metrics can also be written in the warped form of \eqref{rotating ansatz}, their asymptotic behavior differs substantially from that of the boost black string \eqref{boost} and the Kaluza-Klein monopole \eqref{monopole}. Therefore, it is necessary to examine the validity of \eqref{Legendre transformation D} in a separate analysis.

In this section, we demonstrate the validity of our proposed method \eqref{Legendre transformation D} using two examples: the 4D Kerr black hole and arbitrary odd dimension equal angular momenta Kerr black holes.  However, in certain cases, regularization process must be introduced to eliminate divergences, a complication that was absent in the previous section for the boost black string and the Kaluza-Klein monopole.

\subsection{4D Kerr black hole}
We can first consider the Kerr black hole
\bea
d\hat{s}_4^2&=&-\Big(1-\frac{2\mu r}{\Sigma}\Big)dt^2-\frac{4\mu a r}{\Sigma}(1-x^2)dt d\varphi+\frac{\Sigma}{\Delta}dr^2+\Sigma\frac{dx^2}{1-x^2}\cr
&&+\Big(r^2+a^2+\frac{2\mu r}{\Sigma}a^2(1-x^2)
\Big)(1-x^2)d\varphi^2\,,\cr
\Delta&=&r^2-2\mu r+a^2\,,\qquad \Sigma=r^2+a^2x^2\,
\eea
in 4D general relativity
\bea
\hat{S}_4=\frac{1}{16\pi }\int d^4x\sqrt{-\hat{g}} \hat{R}\,.\label{GR 4D}
\eea
Here we set $\cos\theta=x$ and $x\in[-1,1]$. The black hole horizon $r = r_h$ is located at $\Delta(r_h) = 0$. The black hole thermodynamics are
\bea
M&=&\mu\,,\quad T=\frac{r_h-\mu}{2\pi(r_h^2+a^2)}\,,\quad S=\pi(r_h^2+a^2)\,,\cr
J&=&\mu a\,,\quad \Omega_H=\frac{a}{r_h^2+a^2}\,.
\eea
The Euclidean on-shell action is a functional of the temperature $T$ and the angular velocity $\Omega_H$
\bea
F_\mathrm{H}=F(T,\Omega_H)=M-TS-\Omega_H J=\frac{\mu}{2}\,.\label{free energy 4D Kerr}
\eea

Following the previously proposed method, we perform an $S^1$ reduction \eqref{reduction ansatz D}
\bea
d\hat{s}_4^2=e^{-\phi}ds_3^2+e^{\phi}\big( dz+A_{\1}\big)^2\label{4D reduction}
\eea
of the 4D Einstein theory \eqref{GR 4D} to obtain a 3D EMD theory \eqref{EMD D}
\bea
\mathcal{L}_3&=&R-\frac{1}{2}(\nabla\phi)^2-\frac{1}{4}e^{2\phi}F^2\,,
\eea
with the coordinate $\varphi$ identified as the $z$-direction. Thus, the 3D solution takes the form
\bea
ds_3^2&=&-\Delta(1-x^2)dt^2+\big[(r^2+a^2)\Sigma+2\mu a^2 r(1-x^2)\big](1-x^2)\Big(\frac{dr^2}{\Delta}+\frac{dx^2}{1-x^2}\Big)\,,\cr
A_{\1}&=&-\frac{2\mu r a}{(r^2+a^2)\Sigma+2\mu a^2 r(1-x^2)}dt\,,\cr
e^\phi&=&\frac{1-x^2}{\Sigma}\big[(r^2+a^2)\Sigma+2\mu a^2 r(1-x^2)\big]\,.
\eea

According to (\ref{rotating ansatz},\ref{Legendre transformation D}) and the reduction ansatz \eqref{4D reduction}, the total derivative term introduced in the 4D Einstein theory \eqref{GR 4D} is
\bea
\hat{e}^{-1}\hat{\mathcal{L}}_{\mathrm{Legendre}}=d\big(e^{\phi} A_{\1}\wedge\hat{*}\,F_{\2}\big)\,.\label{Legendre 4D Kerr}
\eea

Compared to the previous examples, the on-shell action calculation for rotating black holes involves some subtleties, so we will carefully address them here. Applying Stokes' theorem, the integrand 3-form decomposes into two contributions
\bea
\hat{I}_{\mathrm{Legendre}}&=&-\frac{1}{16\pi}\int \hat{e}^{-1}\hat{\mathcal{L}}_{\mathrm{Legendre}}
=-\frac{1}{16\pi}\oint e^{\phi} A_{\1}\wedge\hat{*}\,F_{\2}\cr
&=&-\frac{1}{16\pi}\int \hat{I}_xd\tau\wedge dx\wedge d\varphi+\hat{I}_rd\tau\wedge dr\wedge d\varphi\,,\cr
\hat{I}_x&=&-\frac{4  \mu ^2a^2 r (1-x^2)}{\Sigma ^2 \big[(r^2+a^2)^2- \Delta a^2 (1-x^2)\big]}\big[
  (r^2-a^2)a^2x^2+r^2 (3 r^2+a^2)\big]\,,\cr
\hat{I}_r&=&\frac{8\mu ^2 a^4  r^2 x (1-x^2)^2}{\Sigma ^2 \big[(r^2+a^2)^2-\Delta a^2   (1-x^2)\big]}\,.
\eea
Following the approach of \cite{Liu:2022wku}, we adopt the integration contour as
\bea
\int _0^\beta d\tau&=&\beta\,,\quad \int_0^{2\pi}d\varphi=2\pi\,,\quad \int _{-1}^{1}dx \hat{I}_x(r=r_h)\,,\quad \int_{r_h}^{\infty}dr \hat{I}_r(x=\pm1)=0\,.
\eea

We find that the contribution from the total derivative term
\bea
\gamma\beta^{-1}\hat{I}_{\mathrm{Legendre}}&=&\gamma\frac{a^2}{2r_h^2}=\gamma\Omega_H J
\eea
 precisely renders the on-shell action a functional of the angular momentum $J$
\bea
F_{\gamma=1}=F(T,J)=M-TS\,.
\eea

\subsection{Rotating black hole in Odd-dimension Einstein Theory}

As the number of spacetime dimensions increases, the number of independent rotation parameters also grows. As a result, higher dimensional rotating black holes feature more rotational degrees of freedom and exhibit a more complex structure than their four-dimensional Kerr counterpart \cite{Myers:1986un,Gibbons:2004js,Gibbons:2004uw}.

Without loss of generality, we focus on a particular simplified case. In odd dimensions, the equal angular momentum Kerr black hole will reduce to a cohomogeneity metric \cite{Myers:1986un,Gibbons:2004js,Gibbons:2004uw}. Following the notation in \cite{Feng:2016dbw}, this solution will be expressed as
\bea
ds^2_{2n+1}&=&-\frac{h(r)}{W(r)}dt^2+\frac{dr^2}{f(r)}+r^2W(r)(\sigma_{n-1}+\omega(r)dt)^2+r^2ds^2_{\mathbb{CP}^{n-1}}\,,\cr
W(r)&=&1+\frac{\nu^2}{r^{D-1}},\quad \omega(r)=\frac{\nu\sqrt{\mu}}{r^{D-1}W(r)},\quad f(r)=h(r)=W(r)-\frac{\mu}{r^{D-3}}\,.\label{rotating metric}
\eea
Here $ds^2_{\mathbb{CP}^{n-1}}$ is a $(n-1)$-dimensions complex projective space and the dimension of spacetimes is $D=2n+1$. Details of the $\mathbb{CP}^m$ manifold are provided in Appendix \ref{CP}.

The thermodynamic are
\bea
M&=&\frac{  (D-2) \Omega _{D-2}}{16 \pi }\mu\,,\quad J=\frac{(D-1) \Omega _{D-2} }{16 \pi }\sqrt{\mu } \nu \,,
\quad \Omega_H=\frac{\nu }{r_{h} \sqrt{r_{h}^{D-1}+\nu ^2}}\,,\cr
T&=&\frac{(D-3) r_{h}^{D-1}-2  \nu ^2}{4 \pi  r_{h}^{\frac{1}{2}(D+1)} \sqrt{r_{h}^{D-1}+\nu ^2}}\,,
\quad S=\frac{\Omega _{D-2}}{4}  r_{h}^{\frac{1}{2}(D-3)} \sqrt{r_{h}^{D-1}+\nu ^2}\,.
\eea
Here $r_{h}$ is the horizon locating at $f(r_{h})=h(r_{h})=0$. As in the case of the 4D Kerr black hole, the on-shell action remains a functional of $(T, \Omega_H)$
\bea
F_\mathrm{H}=F(T,\Omega_H)=M-TS-\Omega_H J.
\eea

Since the metric in equation \eqref{rotating metric} is already expressed in Kaluza-Klein form \eqref{rotating ansatz}, we can directly apply equation \eqref{Legendre transformation D} to carry out the Legendre transformation
\bea
&&\hat{\mathcal{L}}_{\mathrm{Legendre}}=\hat{\nabla}_{\mu}\big(\Omega
\mathcal{F}^{\mu\nu}\mathcal{A}_{\nu}
\big)\,,\cr
&&\Omega=r^2W(r)\,,\quad \mathcal{A}_{\1}=A_{\mathbb{CP}^{n-1}}+\omega(r)dt\,.
\eea
However, we find that due to the presence of $A_{\mathbb{CP}^{n-1}}$, the integral of this total derivative diverges as $r^{D-3}$, a divergence that did not arise in any of the previous cases. Therefore, it is necessary to manually introduce a regularization procedure to remove the divergence. We choose the regularized total derivative term as
\bea
\hat{\mathcal{L}}_{\mathrm{reg}-\mathrm{Legendre}}&=&\hat{\nabla}_{\mu}\left(\Omega
\mathcal{F}^{\mu\nu}(\mathcal{A}_{\nu}-\bar{\mathcal{A}}_{\nu})
\right)\,,\qquad \bar{\mathcal{A}}_{\1}=A_{\mathbb{CP}^{n-1}}\,.
\eea

After regularization, the integral of the total derivative becomes finite and successfully shifts the black hole to a different thermodynamic ensemble
\bea
\beta^{-1}\hat{I}_{\mathrm{Legendre}}&=&-\frac{1}{16\pi\beta}\int d^Dx\sqrt{-\hat{g}} \hat{\mathcal{L}}_{\mathrm{reg}-\mathrm{Legendre}}
=\Omega_H J\,,
\eea
making the on-shell action a functional of $(T, J)$
\bea
F_{\gamma=1}=F(T,J)=M-TS\,.
\eea

\section{Rotating black holes in Chern-Simons theory}\label{FFA}

The five-dimensional minimal supergravity model extends the Einstein-Maxwell theory. Its action is given by the Einstein-Maxwell theory supplemented with a CS term
\bea
\label{supergravity}
\hat{S}=\frac{1}{16\pi }\int d^5x\sqrt{-\hat{g}}\hat{\mathcal{L}}\,,\qquad
\hat{\mathcal{L}}=\hat{R}-\frac{1}{4}\hat{F}^2-\frac{1}{12\sqrt{3}}\epsilon^{\mu\nu\rho\sigma\delta}\hat{F}_{\mu\nu}\hat{F}_{\rho\sigma}\hat{A}_\delta
\eea
or
\bea
\hat{e}^{-1}\hat{\mathcal{L}}=\hat{R}\,\hat{*}\,\oneone-\frac{1}{2}\,\hat{*}\,\hat{F}_{\2}\wedge \hat{F}_{\2}+\frac{1}{3\sqrt{3}}\hat{F}_{\2}\wedge \hat{F}_{\2}\wedge \hat{A}_{\1}\,.\label{supergravity form}
\eea
We choose $\sqrt{-\hat{g}}\epsilon^{01234}=-1$. Unlike the conventional 5D Einstein-Maxwell theory, the presence of the CS term in this model admits rotating charged black hole exact solutions \cite{Cvetic:2004hs,Madden:2004ym,Chong:2005hr}.

It is clear that this theory features the Maxwell potential $\hat{A}_{\1}$ explicitly, in contrast to conventional Maxwell theory, where $\hat{A}_{\1}$ appears only through its field strength $\hat{F}_{\2} = d\hat{A}_{\1}$. If we perform a gauge transformation on the Maxwell field, $\delta_\lambda \hat{A}_{\1} = d\lambda$, we find that the variation of the action \eqref{supergravity form} amounts to a total derivative
\bea
\delta_\lambda\big(\hat{e}^{-1}\hat{\mathcal{L}}\big)=d\big(\frac{\lambda}{3\sqrt{3}}\hat{F}_{\2}\wedge \hat{F}_{\2}\big)\,.\label{CS gauge}
\eea
Therefore, although the EOMs are gauge invariant, this discrepancy between the action and the EOM leads to inconsistencies in defining conserved charges and performing the Legendre transformation of the electric charge and potential.

Therefore, in CS theories, even the Maxwell sector can face consistency issues. Consequently, in addition to the Legendre transformation between angular quantities $(\Omega_H,J)$ in the purely gravitational sector, we must also reconsider the definition of electric charge $Q_e$ in the Maxwell sector, as well as the Legendre transformation for the electric part $(\Phi_e,Q_e)$. This marks a clear departure from the previously discussed cases, where gauge invariance was manifest and such issues did not arise. In this section, working within the framework of dimensional reduction, we use a concrete example to illustrate the issues we encounter and present our proposed resolution. We also use this example to test the reliability of our proposed method.

\subsection{Black hole thermodynamics and the inconsistent electric charge definition}

The EOMs and boundary terms of the minimal supergravity model \eqref{supergravity form} can be obtained by varying the action
\bea
&&\frac{\delta \big(\sqrt{-\hat{g}}\hat{\mathcal{L}}\big)}{\sqrt{-\hat{g}}}=\hat{E}_{\mu\nu}\delta \hat{g}^{\mu\nu}+\hat{S}_A^{\mu}\delta \hat{A}_{\mu}+\hat{\nabla}_\mu\hat{\Theta}_g^\mu+\hat{\nabla}_\mu\hat{\Theta}_A^\mu\,,\cr
&&\hat{E}_{\mu\nu}=\hat{R}_{\mu\nu}-\frac{1}{2}\hat{g}_{\mu\nu}\hat{R}+\frac{1}{8}\hat{g}_{\mu\nu}\hat{F}^2-\frac{1}{2}\hat{F}^2_{\mu\nu}\,,\quad
\hat{\mathbf{S}}_{\hat{A}}=d\Big(-\hat{*}\, \hat{F}_{\2}+\frac{1}{\sqrt{3}}\hat{F}_{\2}\wedge \hat{A}_{\1}\Big)\,,\cr
&&\hat{\Theta}_g^\mu=\hat{g}^{\mu\alpha}\hat{\nabla}^\beta\delta \hat{g}_{\alpha\beta}-\hat{g}^{\alpha\beta}\hat{\nabla}^\mu\delta \hat{g}_{\alpha\beta}\,,\quad
\hat{\mathbf{\Theta}}_{\hat{A}}=\Big(\hat{*}\, \hat{F}_{\2}-\frac{2}{3\sqrt{3}}\hat{F}_{\2}\wedge \hat{A}_{\1}\Big)\wedge\delta \hat{A}_{\1}\,.\label{EOM and surface FFA}
\eea

This theory admits charged black hole solutions \cite{Chong:2005hr} with two independent angular momenta. However, for simplicity, we focus on the case where the angular momenta are equal \cite{Cvetic:2004hs,Madden:2004ym}
\bea
d\hat{s}_5^2&=&-\frac{r^2W(r)}{4b(r)^2}dt^2+\frac{dr^2}{W(r)}+\frac{r^2}{4}\big(\sigma_1^2+\sigma_2^2\big)+b(r)^2\big(\sigma_3+f(r)dt-f(r_h)dt\big)^2\,,\cr
\hat{A}_{\1}&=&\psi_e(r)\Big(dt-\frac{1}{2}j\sigma_3\Big)+cdt\,.\label{FFA black hole solution}
\eea
$c$ is an arbitrary gauge parameter. Here we introduce 1-form $\sigma_i$ on $S^3$
\bea
\sigma_1&=&\cos\chi d\theta+\sin\chi\sin\theta d\varphi\,,\cr
\sigma_2&=&-\sin\chi d\theta+\cos\chi\sin\theta d\varphi\,,\cr
\sigma_3&=&d\chi+\cos\theta d\varphi\,.
\eea
to simplify the metric form. The solution is
\bea
b(r)^2&&=\frac{r^2}{4}  \Big(1+\frac{2 j^2 p}{r^4}-\frac{j^2 q^2}{r^6}\Big)\,,\qquad
f(r)=-\frac{j }{2 b(r)^2}\Big(\frac{2 p-q}{r^2}-\frac{q^2}{r^4}\Big)\,,\cr
W(r)&&=1-\frac{2 (p- q)}{r^2}+\frac{2 j^2 p+q^2}{r^4}\,,\qquad
\psi_e(r)=\frac{\sqrt{3}q}{r^2}\,.
\eea
The black hole horizon locates at $W(r=r_h)=0$.

In the preceding sections, we studied black hole thermodynamics in Einstein-Maxwell theory \eqref{potential and charge} and in EMD theory \eqref{potential and charge EMD} within the Wald formalism. In both cases, the definitions of the electric potential and electric charge are intrinsically encoded in the Wald expressions and coincide with those derived from the equations of motion. By contrast, in the presence of CS terms the situation becomes more subtle.
In Appendix \ref{FFA thermodynamics}, we compute the black hole thermodynamics consistent with the first law. It is worth noting that the electric charge is based on the Maxwell field EOMs \eqref{5D FFA electric part}. However, if we attempt to extract the charge definition directly from the Wald formalism, as done previously, we find that the result does not match that obtained from the EOMs
\bea
\label{Wald electric part}
\delta\hat{\mathbf{Q}}_{\hat{A}}-i_{\hat{\xi}}\hat{\mathbf{\Theta}}_{\hat{A}}&=&\delta\big(\hat{*}\, \hat{F}_{\2}-\frac{2}{3\sqrt{3}}\hat{F}_{\2}\wedge \hat{A}_{\1}\big)\big(-i_{\hat{\xi}} \hat{A}_{\1}\big)\cr
&&-i_{\hat{\xi}}\big(\hat{*}\, \hat{F}_{\2}-\frac{2}{3\sqrt{3}}\hat{F}_{\2}\wedge \hat{A}_{(1)}\big)\wedge\delta \hat{A}_{\1}\,.
\eea
The cause of this ambiguity is the CS theory is not invariant under the gauge transformation \eqref{CS gauge}. When varying the action, both the Maxwell field \(A_{\1}\) and its field strength \(F_{\2}\) contribute to the EOMs \eqref{5D FFA electric part}. However, only \(F_{\2}\) contributes to the surface term \(\mathbf{\Theta}_A\) \eqref{EOM and surface FFA} and the Noether charge \(\mathbf{Q}_A\) \eqref{FFA Noether charge}. This mismatch prevents the direct determination of the electric charge from the Wald formalism.

Some previous works \cite{Gauntlett:1998fz,Rogatko:2006hck,Hanaki:2007mb} argued that $\hat{F}_{\2}\wedge \hat{A}_{\1}$ part decays sufficiently rapidly and hence we can use
\bea
\frac{1}{16\pi}\int_{\infty}\hat{*}\, \hat{F}_{\2}
\eea
to compute electric charges. Although this formula can provide a charge to satisfy the first law and Smarr-type relation, the integrated function $\hat{*}\, \hat{F}_{\2}$ is not a close form in minimal supergravity and therefore this formula is not consistent with the electric charge definition. In \cite{Banados:2005da}, without considering the charge formula, the authors used Wald formalism to work out the first law \eqref{black hole mechanics} and then read the electric charge from it.

This inconsistency affects not only the Wald formalism but also the evaluation of the Euclidean on-shell action. Following the variational procedures for the Euclidean on-shell action in Einstein-Maxwell and EMD theories (\ref{variation}, \ref{Legendre transformation}), we vary the Euclidean on-shell action of the minimal supergravity theory \eqref{supergravity form}, and then substitute the EOMs to simplify, yielding
\bea
\delta\big(\hat{e}^{-1}\hat{\mathcal{L}}\big)=d\Big[\big(\hat{*}\, \hat{F}_{\2}-\frac{2}{3\sqrt{3}}\hat{F}_{\2}\wedge \hat{A}_{\1}\big)\wedge\delta \hat{A}_{\1}
\Big]\,.\label{variation of CS on-shell}
\eea
This appears to be an action consistent with Dirichlet boundary conditions, placing the black hole in an ensemble where the electric potential $\Phi_e$ serves as the thermodynamic variable. Of course this interpretation that is further confirmed by the explicit results
\bea
F_\mathrm{H}=F(T, \Omega_H, \Phi_e)=M-TS-\Omega_HJ-\Phi_eQ_e\,.\label{free energy FFA}
\eea
However, in \eqref{variation of CS on-shell}, the term multiplying \(\delta \hat{A}_{\1}\) does not correspond to the exact definition of the electric charge \eqref{5D FFA electric part}. As a result, it cannot be expressed in terms of thermodynamic quantities in the same form as (\ref{variation impose}, \ref{variation impose EMD}).

Despite these difficulties, it is still possible to construct a self-consistent framework. Drawing inspiration from the Legendre transformation between angular momentum and angular velocity, we observe that for metrics of the form \eqref{rotating ansatz}, one can perform a Kaluza-Klein reduction to obtain a manifestly gauge-invariant action in a lower-dimensional theory, and subsequently examine the consistency of the variational principle. In this section, we demonstrate that for metric ansatze with the specific structure of \eqref{rotating ansatz}, this approach not only provides a consistent definition of electric charge but also allows for a well-defined Legendre transformation between the electric charge $Q_e$ and the electric potential $\Phi_e$, analogous to the case of angular momentum and angular velocity. A similar idea has been proposed in \cite{Elvang:2005sa,Suryanarayana:2007rk}.

\subsection{Kaluza-Klein reduction of the Wald formalism}

From a technical standpoint, the inconsistency between the EOMs and the Wald formalism arises from the fact that every occurrence of $\hat{A}_{\1}$ in the action contributes to the EOMs, whereas only $d\hat{A}_{\1}$ contributes to the Noether charge and the surface term. Therefore, we perform an $S^1$ reduction
\bea
d\hat{s}_5^2&=&e^{-\frac{1}{\sqrt{3}}\phi}ds_4^2+e^{\frac{2}{\sqrt{3}}\phi}\big(dz+\mathcal{A}_{\1}\big)^2,\cr
\hat{A}_{\1}&=&A_{\1}+\psi\big(dz+\mathcal{A}_{\1}\big)\label{KK reduction FFA}
\eea
that all $U(1)$ gauge fields $A_{\1}$ and $\mathcal{A}_{\1}$ appear only through their derivatives. After performing several integrations by parts, the resulting lower-dimensional theory takes the form
\bea
\mathcal{L}&=&R*\oneone-\frac{1}{2}* d\phi\wedge d\phi-\frac{1}{2}e^{\sqrt{3}\phi}*\mathcal{F}_{\2}\wedge\mathcal{F}_{\2}-\frac{1}{2}e^{\frac{1}{\sqrt{3}}\phi}* F_{\2}\wedge F_{\2}\cr
&&-\frac{1}{2}e^{-\frac{2}{\sqrt{3}}\phi}* d\psi\wedge d\psi+\frac{1}{\sqrt{3}}\psi F^0_{\2}\wedge F^0_{\2}+\frac{1}{\sqrt{3}}\psi^2 F^0_{\2}\wedge \mathcal{F}_{\2}+\frac{1}{3\sqrt{3}}\psi^3\mathcal{F}_{\2}\wedge\mathcal{F}_{\2}\,.\label{CS 4D}
\eea
Here we define some shorthand notations
\bea
F_{\2}^0=dA_{\1}\,,\qquad \mathcal{F}_{\2}=d\mathcal{A}_{\1}\,,\qquad F_{\2}=F_{\2}^0+\psi\mathcal{F}_{\2}\,.
\eea
The resulting theory is fully invariant under all $U(1)$ gauge transformations, thereby ensuring consistency between the EOMs
\bea
\mathbf{E}_{\mathcal{A}}&=&d\big(e^{\sqrt{3}\phi}*\mathcal{F}_{\2}+e^{\frac{\phi}{\sqrt{3}}}\psi* F_{\2}-\frac{1}{\sqrt{3}}\psi^2 F^0_{\2}-\frac{2}{3\sqrt{3}}\psi^3\mathcal{F}_{\2}\big)=0\,,\cr
\mathbf{E}_A&=&d\big(e^{\frac{\phi}{\sqrt{3}}}* F_{\2}-\frac{2}{\sqrt{3}}\psi F^0_{\2}-\frac{1}{\sqrt{3}}\psi^2\mathcal{F}_{\2}\big)=0\label{EOM 4 2}
\eea
and Wald formalism
\bea
\mathbf{Q}&=&-* d\xi-\big(e^{\sqrt{3}\phi}*\mathcal{F}_{\2}+e^{\frac{\phi}{\sqrt{3}}}\psi* F_{\2}-\frac{1}{\sqrt{3}}\psi^2 F^0_{\2}-\frac{2}{3\sqrt{3}}\psi^3\mathcal{F}_{\2}\big)\big(i_\xi \mathcal{A}_{\1}\big)\cr
&&-\big(e^{\frac{\phi}{\sqrt{3}}}* F_{\2}-\frac{2}{\sqrt{3}}\psi F^0_{\2}-\frac{1}{\sqrt{3}}\psi^2\mathcal{F}_{\2}\big)\big(i_\xi A_{\1}\big)\,,\cr
\mathbf{\Theta}_{\mathcal{A}}&=&-\big(e^{\sqrt{3}\phi}*\mathcal{F}_{\2}+e^{\frac{\phi}{\sqrt{3}}}\psi* F_{\2}-\frac{1}{\sqrt{3}}\psi^2 F^0_{\2}-\frac{2}{3\sqrt{3}}\psi^3\mathcal{F}_{\2}\big)\wedge\delta\mathcal{A}_{\1}\,,\cr
\mathbf{\Theta}_A&=&-\big(e^{\frac{\phi}{\sqrt{3}}}* F_{\2}-\frac{2}{\sqrt{3}}\psi F^0_{\2}-\frac{1}{\sqrt{3}}\psi^2\mathcal{F}_{\2}\big)\wedge\delta A_{\1}\,.\label{Wald 4 2}
\eea
It is clear that, owing to the manifest \(U(1)\) gauge symmetry of the lower-dimensional theory \eqref{CS 4D}, the Wald formalism \eqref{Wald 4 2} provides a definition of the electric charge that fully agrees with the EOMs \eqref{EOM 4 2}.

We reduce the 5D rotating black hole \eqref{FFA black hole solution} along the $\chi$ coordinate, treating it as the $z$-direction. The resulting four-dimensional black hole solution is:
\bea
ds_4^2&=&\sqrt{b(r)^2}\Big[-\frac{r^2W(r)}{4b(r)^2}dt^2+\frac{dr^2}{W(r)}+\frac{r^2}{4}(\sigma_1^2+\sigma_2^2)\Big]\,,\cr
A_{\1}&=&\big(\psi_e(r)-f(r)\psi(r)\big)dt\,,\quad \psi(r)=-\frac{1}{2}j\psi_e(r)\,,\cr
\mathcal{A}_{\1}&=&fdt+\cos\theta d\varphi\,\qquad \phi=\frac{\sqrt{3}}{2}\log b(r)^2\,.\label{4D black hole}
\eea
This solution automatically satisfies the EOMs \eqref{EOM 4 2} of the four-dimensional theory \eqref{CS 4D}, and its thermodynamics can be consistently derived from the lower-dimensional Wald formalism \eqref{Wald 4 2}.

The equivalence between the two thermodynamic descriptions can be established through a Kaluza-Klein reduction analysis of the Wald formalism. We find that the Wald formalism before and after dimensional reduction \eqref{KK reduction FFA} is equivalent up to a total derivative ambiguity
\bea
&&\delta\hat{\mathbf{Q}}-i_{\hat{\xi}}\hat{\mathbf{\Theta}}=\left(
\delta\mathbf{Q}-i_\xi\mathbf{\Theta}\right)\wedge dz+d\left(\delta\Pi_{Q}+i_\xi\Pi_{\Theta}\right)\wedge dz\,,\cr
&&\Pi_{Q}=-\frac{2}{3\sqrt{3}}\psi A'_{\1}(i_\xi A'_{\1})
+\frac{1}{\sqrt{3}}\psi^2\left(i_\xi A_{\1}\right)  \mathcal{A}_{\1}+\frac{2}{3\sqrt{3}}\psi^3\left(i_\xi\mathcal{A}_{\1}\right) \mathcal{A}_{\1}\,,\cr
&&\Pi_{\Theta}=-\frac{2}{3\sqrt{3}}\psi A'_{\1}\wedge\delta A'_{\1}-\frac{1}{\sqrt{3}}\psi^2\delta A_{\1}\wedge  \mathcal{A}_{\1}-\frac{2}{3\sqrt{3}}\psi^3\delta\mathcal{A}_{\1}\wedge \mathcal{A}_{\1}\,,\cr
&&A'_{\1}= A_{\1}+\psi\mathcal{A}_{\1}\,.
\eea
In fact, as noted in \cite{Ma:2022nwq}, the Wald formalisms of equivalent CS theories can differ by a total derivative ambiguity. This result shows that, although the Wald formalism for CS theory \eqref{Wald electric part} does not formally provide the exact expression for the electric charge, one can always introduce an appropriate total derivative term, exploiting the inherent ambiguity of the formalism, to render it consistent with the charge definition obtained from the EOMs. However, due to the absence of a manifest \(U(1)\) gauge symmetry in CS theory, a strictly covariant expression for this total derivative is not available. In the special case where the black hole geometry possesses a manifest \(U(1)\) isometry \eqref{KK reduction FFA}, this total derivative can be explicitly expressed in the sense of Kaluza-Klein reductions.

Similarly, in 4D, one can readily perform a Legendre transformation between $(Q_A, \Phi_A)$ by introducing an appropriate total derivative term
\bea
e^{-1}\mathcal{L}_A=d\Big[\big(e^{\frac{\phi}{\sqrt{3}}}* F_{\2}-\frac{2}{\sqrt{3}}\psi F^0_{\2}-\frac{1}{\sqrt{3}}\psi^2\mathcal{F}_{\2}\big)\wedge A_{\1}
\Big]\,.
\eea
Returning to 5D, we can implement the Legendre transformation between $(Q_e, \Phi_e)$ for rotating black holes in minimal supergravity by adding the following total derivative term
\bea
&&\hat{e}^{-1}\hat{\mathcal{L}}_{\mathrm{Legendre}}^e=d\big(-\hat{*}\, \hat{F}_{\2}\wedge \hat{A}_{\1}
\big)\,,\cr
\Rightarrow&& \gamma\hat{I}^e=-\frac{\gamma}{16\pi}\int\hat{e}^{-1}\hat{\mathcal{L}}_{\mathrm{Legendre}}^e=\gamma\beta \Phi_eQ_e\,.\label{Legendre FFA ele}
\eea
The on-shell action \eqref{free energy FFA} will change to the functional of electric charge
\bea
F_{\gamma=1}=F(T, \Omega_H, Q_e)=M-TS-\Omega_HJ\,.
\eea
It is clear that, although the exact 5D expression for the electric charge \eqref{5D FFA electric part} was not used in \eqref{Legendre FFA ele}, the integral nonetheless reproduces the Legendre transformation exactly. This result is entirely due to the antisymmetry, and in fact, \eqref{Legendre FFA ele} can equivalently be written
\bea
\hat{e}^{-1}\hat{\mathcal{L}}_{\mathrm{Legendre}}^e=d\Big[-\big(\hat{*}\, \hat{F}_{\2}-m\hat{F}_{\2}\wedge \hat{A}_{\1}\big)\wedge \hat{A}_{\1}
\Big]\,.
\eea
Here, $m$ is an arbitrary number, and it can naturally take values like $\frac{2}{3\sqrt{3}}$ or $\frac{1}{\sqrt{3}}$.

\subsection{The Legendre transformation between $(\Omega_H, J)$}

Returning to the conjugate pair $(\Omega_H, J)$, since the rotation has been reduced via $S^1$ compactification to a lower-dimensional $U(1)$ gauge field $\mathcal{A}_{\1}$, it is natural to treat this pair in the same way as the electric charge and potential
\bea
e^{-1}\mathcal{L}_J&=&d\Big[\big(e^{\sqrt{3}\phi}*\mathcal{F}_{\2}+e^{\frac{1}{\sqrt{3}}\phi}\psi* F_{\2}-\frac{1}{\sqrt{3}}\psi^2 F^0_{\2}-\frac{2}{3\sqrt{3}}\psi^3\mathcal{F}_{\2}\big)\wedge\mathcal{A}_{\1}\Big]\,.
\eea
Of course, we can also employ the method we proposed \eqref{Legendre transformation D} to perform the computation
\bea
&&\hat{e}^{-1}\hat{\mathcal{L}}_{\mathrm{re}-\mathrm{Legendre}}^J=d\Big[(\mathcal{A}_{\1}-\bar{\mathcal{A}}_{\1})\wedge(e^{\frac{2}{\sqrt{3}}\phi}\hat{*}\,\mathcal{F}_{\2})\Big]\,,\cr
&&\mathcal{A}_{\1}=\cos\theta d\varphi+\big(f(r)-f(r_h)\big)dt\,,\qquad\bar{\mathcal{A}}_{\1}=\cos\theta d\varphi\,.
\eea
As in previous cases, it is necessary to manually introduce a regularization procedure to remove the divergence. We find that the result
\bea
\gamma\hat{I}^J=-\frac{\gamma}{16\pi}\int\hat{e}^{-1}\hat{\mathcal{L}}_{\mathrm{re}-\mathrm{Legendre}}^J=\gamma\beta \Omega_HJ\,.
\eea
precisely transforms the on-shell action \eqref{free energy FFA} into a functional of the angular momentum $J$
\bea
F_{\gamma=1}=F(T, J, \Phi_e)=M-TS-\Phi_eQ_e\,.
\eea

\section{More about the $U(1)$ gauge choice in Chern-Simons Euclidean action}\label{More about the $U(1)$ gauge choice in Chern-Simons Euclidean action}

Motivated by the holographic principle \cite{Maldacena:1997re}, there has been growing interest in recent studies of the Euclidean on-shell action \cite{Bobev:2022bjm,Cassani:2022lrk,Ma:2024ynp} for rotating black holes in CS theory \eqref{supergravity} with the cosmological constant term $12/\ell^2$.
However, when computing the on-shell action in minimal supergravity, the choice of gauge for the Maxwell field introduces subtleties. Although \(U(1)\) gauge transformations do not affect the EOMs and thus leave the black hole solution unchanged, they do add a total derivative term to the on-shell action \eqref{CS gauge}. This appears to suggest that the on-shell action of the black hole solution \eqref{FFA black hole solution} in minimal supergravity depends on the choice of gauge for the Maxwell field. Yet, since the on-shell action determines the free energy of the black hole, it cannot depend on the gauge choice. This implies that a specific gauge must be fixed in order to ensure consistency between the black hole on-shell action and its free energy.

While the issue of gauge choice has been addressed in \cite{Guo:2021ikr,Bobev:2022bjm,Ma:2024ynp,Cassani:2022lrk,Cai:2024tyv,Zhao:2025gej}, the conclusions drawn there remain limited in applicability. In this section, we provide a detailed clarification of this problem. We will show that the Maxwell field \(\hat{A}_{\1}\) becomes singular at the black hole's degenerate points, making the CS term ill-defined. That is, while black hole thermodynamic quantities and the free energy are gauge-independent physical observables, a specific gauge choice for the \(U(1)\) field \(\hat{A}_{\1}\) is necessary to render the CS integral well-defined. Only by ensuring that the CS term remains smooth at all degenerate points can the black hole's Euclidean on-shell action be properly defined, thereby yielding a consistent and accurate free energy.

\subsection{Degenerate points in the rotating black hole spacetimes}

As noted in \cite{Guo:2021ikr, Bobev:2022bjm,Ma:2024ynp, Cassani:2022lrk}, it is important to choose a specific gauge for the Maxwell potential $\hat{A}_{\1}$ such that $\hat{A}_\mu \hat{A}^\mu$ remains finite at the horizon $r = r_h$. The gauge parameter $c$ \eqref{FFA black hole solution} is finally chosen as $c = -\Phi_e$, so that $i_{\hat{\xi}} \hat{A}_{\1}$ vanishes at the horizon \cite{Bobev:2022bjm,Cassani:2022lrk,Ma:2024ynp}. The free energy obtained from the Euclidean on-shell action \eqref{free energy FFA} is a functional of $(T, \Omega_H, \Phi_e)$.

Here we demonstrate that the Maxwell field appearing in the CS term may become singular at the degenerate points
\(\{ r = r_h, \infty; \theta = 0, \pi \}\) of the black hole solution \eqref{FFA black hole solution}.
This issue can be diagnosed by examining the behavior of the gauge-invariant scalar
\(\hat{A}_\mu \hat{A}^\mu\) at these locations.
If the Maxwell field fails to remain regular at the degenerate points, the on-shell integral of the CS term is no longer well defined.
To consistently evaluate the CS contribution, it is therefore necessary to fix a specific gauge for \(\hat{A}_\mu\) such that the CS term remains regular at all degenerate points.
The gauge choice \(c = -\Phi_e\), adopted in \cite{Guo:2021ikr, Bobev:2022bjm, Cassani:2022lrk}, precisely guarantees the regularity of the CS term at the horizon \(r = r_h\).
To make this point explicit, we now introduce a more general gauge choice for \(\hat{A}_\mu\)
\bea
\hat{A}_{\1}&=&\psi_e(r)\Big(dt-\frac{1}{2}j\sigma_3\Big)+c_tdt+c_\varphi d\varphi+c_\chi d\chi\,.\label{FFA black hole solution gene}
\eea
From the expression
\bea
\hat{A}_\mu \hat{A}^\mu=-\frac{b^2 }{r^2 W}\big[2 (c_t-f c_{\chi })+(2+f j) \psi _e\big]^2+\frac{4 (c_{\varphi }-\cos\theta c_{\chi
   })^2}{r^2 \sin ^2\theta}+\frac{(j \psi _e-2 c_{\chi })^2}{4 b^2}\,,
\eea
we see that the quantity remains manifestly finite at infinity $r = \infty$. However, at the other three degenerate points, its divergent behaviors are given by
\bea
\hat{A}_\mu \hat{A}^\mu|_{r\rightarrow r_h}&=&-\frac{\big[(c_t+\Phi _e) (r_h^4+j^2 q)+2 j  (r_h^2+q)c_{\chi }\big]^2}{2 r_h^3 (r_h^2+q) (r_h^2-2 j^2-q)}\frac{1}{r-r_h}+\cdots\,,\cr
\hat{A}_\mu \hat{A}^\mu|_{\theta\rightarrow 0}&=&\frac{4 (c_{\chi }-c_{\varphi })^2}{r^2}\frac{1}{\theta^2}+\cdots\,,\cr
\hat{A}_\mu \hat{A}^\mu|_{\theta\rightarrow \pi}&=&\frac{4 (c_{\chi }+c_{\varphi })^2}{r^2}\frac{1}{(\theta-\pi)^2}+\cdots\,.
\eea
To guarantee regularity at all of the above points, the gauge parameters should be chosen as
\bea
c_t=-\Phi_e\,,\qquad c_{\chi }=0\,,\qquad c_{\varphi }=0\,.
\eea
This result serves as a natural generalization of the earlier prescription. Under this gauge choice, the on-shell action accurately captures the black hole free energy \eqref{free energy FFA}. A similar phenomenon also arises in the case of 4D magnetical black holes. However, since the 4D action is manifestly $U(1)$ gauge invariant, the gauge choice discussed above does not affect the evaluation of the on-shell action.

\subsection{Rotating black holes with two independent angular momenta}

To strengthen the generality and robustness of our conclusion, we now turn to a more general setting: black holes in AdS spacetime with two independent angular momenta. The exact solution was constructed in \cite{Chong:2005hr}
\bea
d\hat{s}^2&=&-\frac{\Delta_\theta[(1+\ell^{-2}r^2)\rho^2dt+2Q\nu]dt}{\Xi_a\Xi_b\rho^2}+\frac{2Q\nu\omega}{\rho^2}
+\frac{f}{\rho^4}\Big(\frac{\Delta_\theta dt}{\Xi_a\Xi_b}-\omega\Big)^2\cr
&&+\frac{\rho^2dr^2}{\Delta_r}+\frac{\rho^2d\theta^2}{\Delta_\theta}+\frac{r^2+a^2}{\Xi_a}\sin^2\theta d\varphi^2
+\frac{r^2+b^2}{\Xi_b}\cos^2\theta d\chi^2\,,\cr
\hat{A}_{\1}&=&\frac{\sqrt{3}Q}{\rho^2}\Big(\frac{\Delta_\theta dt}{\Xi_a\Xi_b}-\omega\Big)+c_tdt+c_\varphi d\varphi+c_\chi d\chi\,,\cr
\nu&=&b\sin^2\theta d\varphi+a\cos^2\theta d\chi\,,\qquad \omega=a\sin^2\theta\frac{d\varphi}{\Xi_a}+b\cos^2\theta\frac{d\chi}{\Xi_b}\,,\cr
\Delta_\theta&=&1-a^2\ell^{-2}\cos^2\theta-b^2\ell^{-2}\sin^2\theta\,,\cr
\Delta_r&=&\frac{(r^2+a^2)(r^2+b^2)(1+\ell^{-2}r^2)+Q^2+2abQ}{r^2}-2\mu\,,\cr
\rho^2&=&r^2+a^2\cos^2\theta+b^2\sin^2\theta\,,\qquad \Xi_a=1-a^2\ell^{-2}\,,\qquad \Xi_b=1-b^2\ell^{-2}\,,\cr
f&=&2\mu\rho^2-Q^2+2abQ\ell^{-2}\rho^2\,.\label{CCLP}
\eea
The black hole horizon is located at $r = r_h$, where $\Delta_r(r_h) = 0$.

It is worth noting that in the coordinate system \eqref{CCLP}, the degenerate points in the radial direction remain at $r = r_h, \infty$, whereas the angular degenerate points are located at $\theta = 0, \frac{\pi}{2}$.
 Similarly, at the three degenerate points, the divergent behavior of $\hat{A}_\mu \hat{A}^\mu$ is given by
\bea
&&\hat{A}_\mu \hat{A}^\mu|_{r\rightarrow r_h}=-\frac{(x-\sqrt{3} yQ  r_h^2)^2}{ r_h^4 (a^2 \cos ^2\theta +b^2 \sin ^2\theta +r_h^2)\Delta_r'(r_h)}\frac{1}{r-r_h}+\cdots\cr
&&\hat{A}_\mu \hat{A}^\mu|_{\theta\rightarrow \frac{\pi}{2}}=\frac{\Xi_b c_{\chi }^2}{r^2+b^2}\frac{1}{(\theta-\frac{\pi}{2})^2}+\cdots\,,\cr
&&\hat{A}_\mu \hat{A}^\mu|_{\theta\rightarrow 0}=\frac{\Xi_a c_{\varphi }^2}{r^2+a^2}\frac{1}{\theta^2}+\cdots\,,\cr
&&x=\big[(r_h^2+a^2) (r_h^2+b^2)+a b Q\big] (\Omega _a c_{\varphi }+\Omega _b c_{\chi })\,,\qquad y=\frac{c_t}{\Phi _e}+1\,.
\eea
Here, the divergent behaviors are rewritten in terms of the electric potential $\Phi_e$ and the two independent angular velocities $\Omega_{a,b}$
\bea
\Phi_{e}&=&\frac{\sqrt{3}Qr_h^2}{abQ+(r_h^2+a^2)(r_h^2+b^2)}\,,\quad
\Omega_{a}=\frac{a(r_h^2+b^2)(1+\ell^{-2}r_h^2)+bQ}{abQ+(r_h^2+a^2)(r_h^2+b^2)}\,,\quad
\Omega_{b}=\Omega_{a}|_{a\leftrightarrow b}\,
\eea
and other thermodynamic quantities have been fully derived using standard methods in \cite{Chong:2005hr}.

To ensure that $\hat{A}_\mu \hat{A}^\mu$ remains regular at all degenerate points, the gauge parameter must be chosen as
\bea
c_t=-\Phi _e\,,\qquad c_{\varphi }=0\,,\qquad c_{\chi }=0\,.
\eea

Under this gauge choice, it can be verified that the Euclidean on-shell action correctly yields the free energy of the black hole solution \eqref{CCLP}. It is important to note that when using the counterterm method to compute the action, a nontrivial Casimir energy appears for $a \neq b$ \cite{Papadimitriou:2005ii}
\bea
\beta^{-1}I=F(T, \Omega_a, \Omega_b, \Phi_e)+\frac{3 \pi  \ell ^2}{32}\Big(1+\frac{(\Xi _a-\Xi _b)^2}{9 \Xi _a \Xi _b}\Big)
\eea
and the proper treatment of this term has been provided in \cite{Gibbons:2005jd}.



\section{Conclusions}\label{Conclusions}

In this paper, we clarify how to properly implement Legendre transformations of black hole free energies from the viewpoint of the Euclidean on-shell action and the associated boundary conditions. To enhance the reliability of our analysis, we adopt the covariant phase space approach to define black hole physical quantities in a precise and systematic manner. This ensures that the black hole free energy derived from the Euclidean on-shell action is fully consistent with the first law of black hole mechanics.

For 4D Maxwell theory, we provided precise definitions of the electric sector $(Q_e, \Phi_e)$ and the magnetic sector $(Q_m, \Phi_m)$ using the Wald formalism and electromagnetic duality. The variational principle clarifies that different boundary conditions for the Maxwell field $A_\mu$ correspond to different thermodynamic potentials. By introducing appropriate total derivative terms, one can shift between these boundary conditions and perform Legendre transformations. However, unlike in conventional thermodynamics, we find that for dyonic black holes, only two ensembles are consistent with the variational principle: those that fix $(Q_e, \Phi_m)$ or $(Q_m, \Phi_e)$. No other ensemble satisfies the consistency conditions. This conclusion is of central importance. Many black hole studies, such as analyses of thermodynamic stability, must be formulated within specific thermodynamic ensembles. If black hole thermodynamics were treated on the same footing as ordinary thermodynamic systems, one could freely perform Legendre transformations among any thermodynamically conjugate variables. Our result, however, shows that the black hole free energy defined through the Euclidean on-shell action is fundamentally different: thermodynamic ensembles incompatible with the imposed boundary conditions cannot be consistently realized. Consequently, when investigating problems such as the thermodynamic stability of dyonic black holes, one must first ensure that the chosen ensemble is compatible with the boundary conditions of the on-shell action. Otherwise, one risks carrying out an ill-defined analysis in a thermodynamic ensemble in which the dyonic black hole does not exist.

Inspired by the Legendre transformation in the Maxwell sector, we turn our attention to physical quantities of purely geometric origin, such as angular momentum and angular velocity. Although the highly nonlinear nature of gravity makes it difficult to introduce total derivative terms for a Legendre transformation in the same straightforward manner as for Maxwell fields, a dimensional reduction perspective provides a natural resolution: Because a rotating black hole admits the Killing vector \(\partial_\varphi\), the spacetime metric necessarily exhibits a \(U(1)\) isometry along the azimuthal direction. From a lower-dimensional viewpoint, higher-dimensional rotation is equivalently described as a \(U(1)\) Maxwell field generated via Kaluza-Klein reduction.
 In this sense, performing a Legendre transformation between $(\Omega_H, J)$ is not only justified but also technically feasible. Furthermore, for metrics with a $U(1)$ fiber bundle structure, we are able to explicitly construct the total derivative term required for such a transformation.

The same approach extends naturally to CS theories. Using five-dimensional minimal supergravity as an example, we demonstrated that although the non-covariance of the action complicates both the definition of conserved charges and the implementation of Legendre transformations, these difficulties can be overcome when the metric admits a $U(1)$ fiber bundle structure. In such cases, performing a Kaluza-Klein reduction leads to a manifestly gauge-invariant lower-dimensional action, from which one can consistently define charges and carry out Legendre transformations in a manner compatible with both the EOMs and the Wald formalism.

Similarly, we find that, due to the lack of manifest $U(1)$ gauge invariance, the Maxwell field in CS theory is significantly more sensitive to its behavior at degenerate points than in normal Maxwell theory. To obtain a well-defined on-shell action, it is necessary to fix a specific gauge for the Maxwell field so as to remove its singularities at all degenerate points.

For researchers investigating the thermodynamic stability of dyonic black holes, performing a Legendre transformation between angular momentum and angular velocity is fully justified, as demonstrated in this work. However, special care must be taken in the electromagnetic sector: fixing both electric and magnetic charges simultaneously is not allowed, since such an ensemble contradicts the variational principle. For Chern-Simons theories without manifest gauge symmetry or covariance, the formulation of a well-defined on-shell action remains a subtle issue requiring further investigation. In this work, we have focused on the simplest five-dimensional minimal supergravity model, where the Chern-Simons term involves only the pure Maxwell sector. Once \(\alpha'\) corrections are included, however, mixed Chern-Simons terms coupling the gravitational and Maxwell fields generically appear. Determining an appropriate gauge choice for such more general Chern-Simons terms is an important open problem that warrants future study.

\section*{Acknowledgement}

We are grateful to Pengju Hu, Hong Lu, Yi Pang, Yanqi Wang and Junkun Zhao for useful discussions. L.M.~is supported in part by the National Key R\&D Program No.~2022YFE0134300, National Natural Science Foundation of China (NSFC) grant No.~12447138 and No.~11935009, Postdoctoral Fellowship Program of CPSF Grant No.~GZC20241211 and the China Postdoctoral Science Foundation under Grant No.~2024M762338.

\appendix

\section{A brief introduction to $\mathbb{CP}^m$}\label{CP}
For rotating black holes in arbitrary odd dimensions with equal angular momenta \eqref{rotating metric}, the spacetime geometry can be represented as a \(U(1)\) fiber bundle over the \(\mathbb{CP}^m\) manifold. In this coordinate system, rotation occurs solely along the fiber direction, while the \(\mathbb{CP}^m\) base manifold remains non-rotating. Here, we provide a brief review of projective \(\mathbb{CP}\) space. Follow the convention in \cite{Cvetic:2018ipt}, the $2m$-dimension $\mathbb{CP}^m$ metric can be expressed by $\mathbb{CP}^{m-1}$
\bea
\label{CPm}
ds^2_{\mathbb{CP}^m}=d\xi^2_m+\sin^2{\xi_m}\cos^2{\xi_m}\sigma_{m-1}^2+\sin^2{\xi_m}ds^2_{\mathbb{CP}^{m-1}}\,,\quad \sigma_{m-1}=d\psi_m+A_{\mathbb{CP}^{m-1}}\,.
\eea
Here $J_{\mathbb{CP}^m}=\frac{1}{2}d\sigma_m=\frac{1}{2}dA_{\mathbb{CP}^m}$ is the K$\ddot{a}$hler form on $\mathbb{CP}^m$. The 1-form potential $A_{\mathbb{CP}^m}$ is given by
\bea
\label{1-form}
A_{\mathbb{CP}^m}=\sin^2{\xi_m}(d\psi_m+A_{\mathbb{CP}^{m-1}})\,,\quad A_{\mathbb{CP}^1}=\sin^2{\xi_1}d\psi_1\,.
\eea
For convenient, we often set $\sin{\xi_m}=x_m,\ \psi_m=y_m$ and then
\bea
\label{CPm easy}
&&ds^2_{\mathbb{CP}^1}=\frac{dx_1^2}{1-x_1^2}+x_1^2(1-x_1^2)dy_1^2\,,\quad A_{\mathbb{CP}^1}=x_1^2dy_1\cr
&&ds^2_{\mathbb{CP}^m}=\frac{dx_m^2}{1-x_m^2}+x_m^2(1-x_m^2)\sigma_m^2+x_m^2ds^2_{\mathbb{CP}^{m-1}}\,,\quad A_{\mathbb{CP}^m}=x_m^2(dy_m+A_{\mathbb{CP}^{m-1}})\,.
\eea

As is well known, every odd dimension sphere $S^{2m+1}$ can be expressed as a $U(1)$ bundle over $\mathbb{CP}^m$
\bea
d\Omega^2_{2m+1}=\sigma_m^2+ds_{\mathbb{CP}^m}^2\,.
\eea
It's easy to see $\mathbb{CP}^1$ is a 2-sphere $S^2$ from (\ref{CPm},\ref{CPm easy}).

\section{Thermodynamic for 5D minimal supergravity and its lower dimension theory}\label{FFA thermodynamics}

In this section, we present the thermodynamics of black holes in minimal supergravity and its lower-dimensional reduced theories.

\subsection{5D rotating charge black hole}
The Killing vector $\hat{\xi}$ of the rotating metric \eqref{FFA black hole solution}
\bea
\hat{\xi}=\partial_t+\Omega_H\partial_\varphi\,,\quad \Omega_H=\frac{2 j (r_h^2+q)}{r_h^4+j^2 q}\label{Killing vector FFA}
\eea
becomes null at horizon $\hat{\xi}^2|_{r=r_h}=0$. The Killing vector generates conserved charges corresponding to the mass and angular momentum. To obtain thermodynamic quantities consistent with the first law, we employ the Wald formalism. The Noether charge in Wald formalism is
\bea
\hat{\mathbf{Q}}&=&\hat{\mathbf{Q}}_g+\hat{\mathbf{Q}}_{\hat{A}}\,,\cr
\hat{\mathbf{Q}}_g&=&-\hat{*}\,d\hat{\xi}\,,\qquad \hat{\mathbf{Q}}_{\hat{A}}=-\big(\hat{*}\, \hat{F}_{\2}-\frac{2}{3\sqrt{3}}\hat{F}_{\2}\wedge \hat{A}_{\1}\big)(i_{\hat{\xi}} \hat{A}_{\1})\,.\label{FFA Noether charge}
\eea
Together with the surface terms \eqref{EOM and surface FFA}, we can compute the mass $M$ and angular momentum $J$ consistently
\bea
\label{Wald formalism}
\delta \hat{\mathcal{H}}&=&\frac{1}{16\pi}\int_{S^3}\big( \delta\hat{\mathbf{Q}}-i_{\hat{\xi}}\hat{\mathbf{\Theta}}\big)=\delta M-\Omega_H\delta J\,,\cr
M&=&\frac{3\omega_3}{8\pi} (p-q)\,,\quad J=\frac{\omega_3}{8\pi} j (2 p-q)\,.\label{5D FFA M and J}
\eea
The temperature is defined in terms of the surface gravity $\kappa$, which is computed from the Killing vector
\bea
\kappa^2=-\frac{\hat{g}^{\mu\nu}\partial_\mu\hat{\xi}^2\partial_\nu\hat{\xi}^2}{4\hat{\xi}^2}\Big|_{r=r_h}\,,\quad T=\frac{\kappa}{2\pi}=\frac{1}{2\pi}\frac{(r_h^2-2j^2-q)(r_h^2+q)}{\sqrt{r_h^2-j^2}(r_h^4+j^2q)}.
\eea
The black hole entropy is computed using Iyer-Wald formula
\bea
S=-\frac1{8}\int_{S^3} d\Omega_3\frac{\partial \hat{\mathcal{L}}}{\sqrt{-\hat{g}}\partial \hat{R}_{\mu\nu\rho\sigma}}\epsilon_{\mu\nu}\epsilon_{\rho\sigma}\bigg|_{r=r_h}
=\frac{\omega_3}{4}\frac{r_h^4+j^2q}{\sqrt{r_h^2-j^2}}\,.
\eea
Here, we define and compute the electric charge based on the EOM for the Maxwell field \eqref{EOM and surface FFA}
\bea
Q_e&=&\frac{1}{16\pi}\int_{S^3}\big(\hat{*}\,\hat{F}_{(2)}-\frac{1}{\sqrt{3}}\hat{F}_{\2}\wedge \hat{A}_{\1}\big)=\frac{\sqrt{3}\omega_3}{8\pi} q,\cr
\Phi_e&=&i_{\hat{\xi}} \hat{A}_{\1}|_{r=r_h}^{r\rightarrow\infty}=\frac{\sqrt{3} q (r_h^2-j^2)}{r_h^4+j^2 q}
\label{5D FFA electric part}
\eea
and all the thermodynamic satisfy the first law
\bea
\delta M=T\delta S+\Phi_e\delta Q_e+\Omega_H\delta J\,.\label{black hole mechanics}
\eea

\subsection{Reduced 4D black hole}

After the $S^1$ reduction in \eqref{KK reduction FFA}, the resulting 4D black hole \eqref{4D black hole} becomes static, with the Killing vector given by $\xi = \partial_t$.

From the 4D perspective, the rotation of the 5D black hole manifests as a charge under a $U(1)$ gauge field $\mathcal{A}_{\1}$. As a result, its associated charge and potential can be computed using standard methods.
\bea
\Phi_A&=&i_\xi A_{\1}|_{r=r_h}^{r\rightarrow\infty}=\frac{\sqrt{3} q (r_h^2-j^2)}{r_h^4+j^2 q}\,,\cr
\Phi_{\mathcal{A}}&=&i_\xi \mathcal{A}_{\1}|_{r=r_h}^{r\rightarrow\infty}=\frac{2 j (r_h^2+q)}{r_h^4+j^2 q}
\eea
We find that the electric potential $\Phi_e$ of the original 5D Maxwell field \eqref{5D FFA electric part} remains consistent with its 4D counterpart $\Phi_A$, while the black hole's angular velocity $\Omega_H$ in \eqref{Killing vector FFA} is recast as the potential $\Phi_{\mathcal{A}}$ associated with the emergent $U(1)$ gauge field $\mathcal{A}_{\1}$.

Because the action is manifestly gauge invariant, the charges obtained from the EOMs \eqref{EOM 4 2} are in complete agreement with those derived from the Wald formalism \eqref{Wald 4 2}
\bea
Q_A=\frac{\sqrt{3}\omega_2}{64\pi} q\,,\qquad
Q_\mathcal{A}=\frac{\omega_2}{64\pi} j (2 p-q).
\eea
Comparing the 5D version (\ref{5D FFA M and J},\ref{5D FFA electric part}), the difference arises from the volume of $d\chi$
\bea
\int_0^{4\pi}d\chi=\frac{8\omega_3}{\omega_2}\,.
\eea
All other thermodynamic calculations proceed conventionally, with the only difference from the 5D case being the factor mentioned above
\bea
M&=&\frac{3\omega_2}{64\pi} (p-q)\,,\qquad T=\frac{1}{2\pi}\frac{(r_h^2-2j^2-q)(r_h^2+q)}{\sqrt{r_h^2-j^2}(r_h^4+j^2q)}\,,\qquad
S=\frac{\omega_2}{32}\frac{r_h^4+j^2q}{\sqrt{r_h^2-j^2}}\,
\eea
and they satisfy the first law
\bea
\delta M=T\delta S+\Phi_A\delta Q_A+\Phi_{\mathcal{A}}\delta Q_{\mathcal{A}}\,.
\eea

\end{document}